\documentclass[onecollarge]{svjour2}       % onecolumn "king-size"
\smartqed  % flush right qed marks, e.g. at end of proof
\usepackage{amsfonts }
\usepackage{bbm}
\usepackage[dvipdfmx]{graphicx}
\usepackage{amsmath,amssymb}
\usepackage{mathrsfs}%hanamoji
\usepackage{dcolumn}% Align table columns on decimal point
\usepackage{bm}% bold math
\usepackage{tabularx}
\usepackage{multirow}
\usepackage{color}
\usepackage{ascmac}

\usepackage{times}
\usepackage{comment}
\usepackage{ulem}
\usepackage{footnote}

\newcommand{\tb}{\textbf}

\newcommand{\ra}{\rangle}
\newcommand{\la}{\langle}

\newcommand{\eps}{\epsilon}

\newcommand{\TR}{{\rm TR}}

\newcommand{\cut}{{\rm cut}}

\newcommand{\ID}{\mathfrak{i}}
\newcommand{\erf}{{\rm erf}}

\def\XXint#1#2#3{{\setbox0=\hbox{$#1{#2#3}{\int}$}
\vcenter{\hbox{$#2#3$}}\kern-.5\wd0}}

\journalname{Journal of Statistical Physics}
\begin{document}

\title{Exact solution to a generalised Lillo-Mike-Farmer model with heterogeneous order-splitting strategies}


\author{Yuki Sato \and Kiyoshi Kanazawa}


\institute{
			\email{sato.yuki.22r@st.kyoto-u.ac.jp}\\
			\email{kiyoshi@scphys.kyoto-u.ac.jp}\\
			Department of Physics, Graduate School of Science, Kyoto University, Kyoto 606-8502, Japan
}

\date{Received: date / Accepted: date}

\maketitle

% OK
\begin{abstract}
	The Lillo-Mike-Farmer (LMF) model is an established econophysics model describing the order-splitting behaviour of institutional investors in financial markets. In the original article (LMF, Physical Review E \tb{71}, 066122 (2005)), LMF assumed the homogeneity of the traders' order-splitting strategy and derived a power-law asymptotic solution to the order-sign autocorrelation function (ACF) based on several heuristic reasonings. This report proposes a generalised LMF model by incorporating the heterogeneity of traders' order-splitting behaviour that is exactly solved without heuristics. We find that the power-law exponent in the order-sign ACF is robust for arbitrary heterogeneous intensity distributions. On the other hand, the prefactor in the ACF is very sensitive to heterogeneity in trading strategies and is shown to be systematically underestimated in the original homogeneous LMF model. Our work highlights that the ACF prefactor should be more carefully interpreted than the ACF power-law exponent in data analyses.
\end{abstract}

\section{Introduction}
	% OK
	Market microstructure of financial markets has been studied quantitatively and empirically in  econophysics~\cite{StanleyB,SlaninaB,BouchaudB,SocialPhysics}. Econophysicists propose various dynamical models, such as at the limit-order book level (e.g., the Santa Fe model~\cite{ZIOB-PRL2003,ZIOB-QFin2003,ZIOB-Bouchaud2002}, the $\epsilon$-intelligence model~\cite{TothPRX2011}, and the latent order-book model~\cite{Donier2015}) and the individual-traders level (e.g., the dealer model~\cite{Takayasu1992,Sato1998,KanazawaPRL2018,KanazawaPRE2018,KanazawaJSP2023}) with the hope that the statistical-physics program is useful even for financial modelling. This paper focuses on a microscopic model of market-order submissions proposed by Lillo, Mike, and Farmer (LMF) in 2005~\cite{LMF_PRE2005}, which was hypothetically based on the order-splitting behaviour of individual traders.

	% OK
	The LMF model is a stylised dynamical model to explain the persistence of market-order flows. In financial data analyses, the binary order-sign sequence of market-order flows is known to be predictable for a long time: by writing a buy (sell) order at time $t$ as $\eps_t=+1$ ($\eps_t=-1$), the autocorrelation function (ACF) of the order-sign sequence obeys the slow decay characterised by the power law, such that 
	\begin{equation}
		C_\tau := \la \eps_t\eps_{t+\tau}\ra \simeq c_0 \tau^{-\gamma} \>\>\> \mbox{for large }\tau, \>\>\> \gamma \in (0,1).
		\label{eq:LRC_ACF}
	\end{equation}
	Here the ensemble average of any stochastic variable $A$ is denoted by $\la A\ra$, $c_0$ is the ACF prefactor, and $\gamma$ is the ACF power-law exponent. This slow decay is called the {\it long-range correlation} (LRC) of the order flows and has been under debate in econophysics and market microstructure for a long time~\cite{BouchaudB}. For example, some researchers state that the LRC is a consequence of herding among traders~\cite{LeBaronPhysA2007,LeBaronEEJ2008,YamamotoJEDC2011}. However, from the viewpoint of empirical support, the current most promising microscopic hypothesis is the order-splitting hypothesis stating that the LRC originates from the order-splitting behaviour of institutional investors. The LMF model is based on this order-splitting hypothesis in describing the LRC from the microscopic dynamics in the spirit of the statistical-physics programs.

	% OK
	The order-splitting hypothesis states that there are traders who split large metaorders into a long sequence of child orders. Because all the child orders share the same sign for a while, the LRC naturally appears in this scenario. The LMF model is a simple stochastic model implementing this order-splitting picture. In the original article~\cite{LMF_PRE2005}, they made the following assumptions: 
	\begin{itemize}
		\item There are $M$ traders in the financial markets. $M$ is a time constant (i.e., a closed system). 
		\item All traders are order-splitting traders, and the homogeneity of their strategy is assumed. 
		\item The distribution of metaorder length $L$ is given by the discrete parato distribution ($L=1,2,...$), such that 
					\begin{equation}
						\rho(L) \simeq \alpha L^{-\alpha-1} \>\>\> \mbox{with } \>\>\alpha \in (1,2).
					\end{equation}
		\item They randomly submit market orders with the same intensity.  
	\end{itemize}
	While this microscopic dynamics is described as an $2M+1$-dimensional stochastic process, LMF solved this model to study the LRC in the ACF as its macroscopic dynamical behaviour with heuristic but reasonable approximations. They finally showed that the ACF asymptotically obeys the LRC asymptotics~\eqref{eq:LRC_ACF}, and the power-law exponent $\gamma$ and the prefactor $c_0$ are given by 
	\begin{subequations}
		\label{eq:LMF_original_LRC}
	\begin{align}
		\gamma &= \alpha -1,  \label{eq:LMF_original_powerlaw_LRC}\\
		c_0 &= \frac{1}{\alpha M^{2-\alpha}}. \label{eq:LMF_original_prefactor_LRC}
	\end{align}		
	\end{subequations}
	They also numerically showed that the power-law exponent formula~\eqref{eq:LMF_original_powerlaw_LRC} robustly works even for an open-system version, where the total number of the traders $M$ fluctuates in time. Since the predictive formula~\eqref{eq:LMF_original_LRC} connects the quantitative relationship between the macroscopic LRC phenomenon and the microscopic parameters, the LMF theory belongs to typical statistical-physics programs and is exceptionally appealing to econophysicists theoretically.

	% OK
	Several empirical studies support both the order-splitting hypothesis and the LMF model. While the original LMF paper could not establish their prediction~\eqref{eq:LMF_original_powerlaw_LRC} at a quantitative level\footnote{They showed that the theoretical line~\eqref{eq:LMF_original_powerlaw_LRC} passed through the centre of the scatterplot between $\alpha$ and $\gamma$, but the regression coefficient did not agree with their prediction at all. In our view, this might be partially due to their imperfect dataset and statistical analyses at that time.} due to the data unavailability of high-quality microscopic datasets in 2005, T\'oth et al. showed very convincing qualitative evidence in 2015 that the order-splitting is the main cause of the LRC by decomposing the total ACF~\cite{TothJEDC2015}. Furthermore, Sato and Kanazawa showed crucial evidence in 2023 that the LMF prediction~\eqref{eq:LMF_original_powerlaw_LRC} precisely works well even at a quantitative level~\cite{SatoPRL2023,SatoPRR2023} using a large microscopic dataset of the Tokyo Stock Exchange (TSE) market. 

	% OK
	At the same time, theoretically, the predictive power of the LMF model is expected to be limited regarding the prefactor $c_0$ because the formula~\eqref{eq:LMF_original_prefactor_LRC} should depend on system-specific details of the underlying microscopic dynamics. Indeed, we noticed that heterogeneity of order-splitting strategies is present during the data analyses for Ref.~\cite{SatoPRL2023,SatoPRR2023} and that such heterogeneity can impact the prefactor $c_0$, while the power-law formula~\eqref{eq:LMF_original_powerlaw_LRC} robustly holds. Given the recent breakthrough in data analyses, we believe the classical LMF theory can be updated to take into account the heterogeneity in trading strategies toward precise data calibration. 

	% OK
	In this report, we propose a generalised LMF model by incorporating heterogeneity of order-splitting strategies. In addition, we solve the generalised LMF model exactly to show the following two characters: (i) The power-law exponent formula~\eqref{eq:LMF_original_powerlaw_LRC} robustly holds true even in the presence of heterogeneous intensity distributions. (ii) The prefactor formula~\eqref{eq:LMF_original_prefactor_LRC} is replaced with a new formula that is sensitive to the intensity distribution. (iii) Furthermore, the classical prefactor formula~\eqref{eq:LMF_original_prefactor_LRC} systematically underestimates the actual prefactor in the presence of heterogeneity in agents. Our results imply that while the interpretation of the ACF power-law exponent is robust and straightforward, the interpretation of the ACF prefactor needs more careful investigation for data calibrations. 

	% OK
	This report is organised as follows. Section~\ref{sec:model} describes our model and mathematical notation with the assumption of the closed system. We show the exact solution for the generalised LMF model in Sec.~\ref{sec:exactsol}. In Sec.~\ref{sec:example}, we study several specific but important cases with numerical verifications. Section~\ref{sec:discussion} discusses the implication of our heterogeneous LMF formulas for realistic data calibration. We conclude our paper with some remarks in Sec.~\ref{sec:conclusion}. At the end of this report, eight appendices follow the main text for its supplements. 

\section{Model}\label{sec:model}
	In this section, let us define the stochastic dynamics of our generalised LMF model. 
	
	\subsection{Mathematical notation}
	% OK
		In this report, the probability density function (PDF) of a stochastic variable $A$ is written as $P(A)$. If the stochastic variable explicitly depends on time $t$, such that $A_t$, the PDF of $A_t$ is denoted by $P_t(A)$. We note that any PDF must satisfy the normalisation condition $\sum_A P(A) = 1$. 
		We also define the cumulative distribution function (CDF) and the complementary cumulative distribution function (CCDF) by 
		\begin{equation}
			P_<(A) := \sum_{A'< A} P(A'), \>\>\> P_{\geq}(A) := \sum_{A'\geq A} P(A') = 1 - \sum_{A' < A}P(A'),
		\end{equation}
		respectively. 
		The stationary PDF and the stationary ensemble average are respectively defined by 
		\begin{equation}
			P_{\rm st} (A) := \lim_{t\to \infty}P_t(A), \>\>\> 
			\la A_t\ra_{\rm st} := \lim_{t\to \infty}\la A_t\ra = \sum_A AP_{\rm st}(A).
		\end{equation}
		Also, under the condition $B$, the conditional PDF and conditional average of $A$ are respectively defined by 
		\begin{equation}
			P(A|B) := \frac{P(A,B)}{P(B)}, \>\>\>
			\la A | B\ra = \sum_{A'} A' P(A'|B).
		\end{equation}

	\subsection{Model parameters and variables}
		\begin{figure}
			\centering
			\includegraphics[width=150mm]{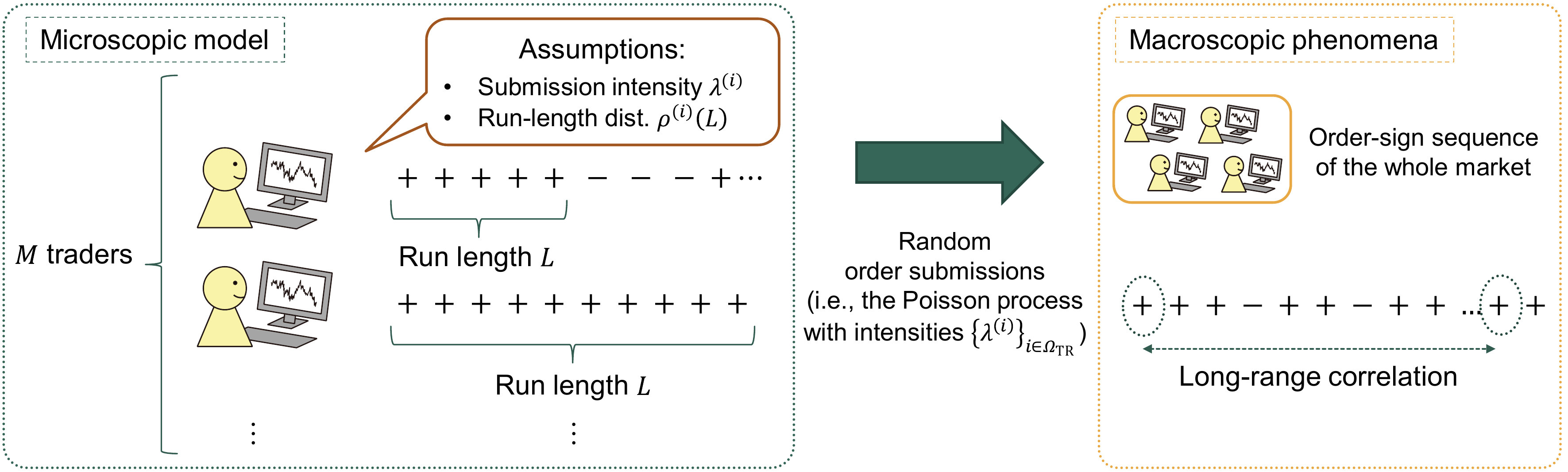}
			\caption{
				Schematic of our generalised LMF model. The total number of the traders is $M:=|\Omega_{\rm TR}|$, which is a time constant positive integer. Any trader $i$ is characterised by the intensity $\lambda^{(i)}$ and the run-length (metaorder-length) distribution $\rho^{(i)}(L)$. Here, a run length is defined by the number of successively same order signs (e.g., $L=3$ for $+++$), and is called the metaorder length in this paper. The intensities and the metaorder-length distribution must satisfy the normalisation conditions $\sum_{i'\in \Omega_{\rm TR}}\lambda^{(i')}=1$ and $\sum_{L=1}^\infty \rho^{(i)}(L)=1$ for any $i\in \Omega_{\rm TR}$. At each timestep, a trader $\ID_{t}$ is randomly selected according to the probability distribution $\{\lambda^{(i)}\}_{i\in \Omega_{\rm TR}}$ (i.e., the discrete-time Poisson process), and then submits a market order.
			}
			\label{fig:setup_LMF}
		\end{figure}
		$\Omega_\TR$ denotes the set of all the traders, and the system is assumed to be closed, such that $M:= |\Omega_\TR|={\rm const}$ (see Fig.~\ref{fig:setup_LMF}). $|\Omega_\TR|$ is a positive integer, and the traders set $\Omega_\TR$ can be written as 
		\begin{equation}
			\Omega_\TR = \{1,2,...,M\}
		\end{equation}
		without loss of generality. We incorporate the heterogeneity of trading strategies into our model, and the characteristic parameters of the $i$th trader are given by the submission rate $\lambda^{(i)}$ and metaorder-length (or run-length) distribution $\rho^{(i)}(L)$. For simplicity, we assume that the executed volume size is always the minimum unit of transactions. In other words, our model is completely characterised by the following parameter set 
		\begin{equation}
			\mathcal{P} := 
			\left(M, \{\lambda^{(i)}\}_{i\in \Omega_{\TR}}, \{\rho^{(i)}(L)\}_{i\in \Omega_{\TR}}\right),
		\end{equation}
		where the submission rate and the metaorder-length distribution satisfy the normalisation of the probability
		\begin{equation}
			\sum_{i'\in\Omega_{\TR}} \lambda^{(i')}=1, \>\>\>
				\sum_{L=1}^{\infty} \rho^{(i)}(L)=1 
		\end{equation} 
		for any $i \in \Omega_{\TR}$.

	% OK
	We next define the state variable of the $i$th trader. The trader $i$ has two state variables $\eps^{(i)}_t$ and $R^{(i)}_t$, representing the order-sign of the metaorder ($\eps^{(i)}_t=+1$ denotes buy and $\eps^{(i)}_t=-1$ denotes sell) and the remaining metaorder length, respectively. The order-sign of the whole market is denoted by $\eps_t$. Thus, this system is specified by the point in the phase space
		\begin{equation}
			X_t: = \left(\eps_t; \eps_t^{(1)}, R_t^{(1)}; \dots; \eps_t^{(M)}, R_t^{(M)}\right)
		\end{equation}
		and is designed as a Markovian stochastic process with dimension $2M+1$.

	\subsection{Stochastic dynamics}
		%OK
		We next proceed with the definition of the stochastic dynamics. Let $\ID_t$ be the stochastic variable representing the trader identifier (ID) who submits the market order at time $t$, such that $\ID_{t} \in \Omega_{\TR}$. We assume that $\ID_{t+1}$ obeys the PDF $\{\lambda^{(\ID)}\}_{\ID \in \Omega_{\TR}}$. In other words, the probability $\ID_{t+1}$ is given by 
		\begin{subequations}
			\label{eq:Dynamics}
			\begin{equation}
				P_{t+1}(\ID) = \lambda^{(\ID)}
			\end{equation}
			as an independent and identically distributed sequence $\{\ID_t\}_{t}$. After the execution by the trader $\ID_{t+1}$, the remaining volume $R^{(\ID_{t+1})}_{t+1}$ decreases by one if $R^{(\ID_{t+1})}_{t} > 1$. If all the metaorder is executed at time $t+1$ (i.e., $R^{(\ID_{t+1})}_t=1$), the metaorder length and its sign are randomly reset for the trader $\ID_{t+1}$. In summary, the dynamics of $X_t$ is given as follows for all $i \in \Omega_{\TR}$ (see Fig.~\ref{fig:setup_LMF} for a schematic): 
			\begin{align}
				R^{(i)}_{t+1} &= 
				\begin{cases}
					R^{(i)}_t & \mbox{if $i \neq \ID_{t+1}$} \\
					R^{(i)}_t - 1 & \mbox{if $i = \ID_{t+1}$ and $R^{(i)}_{t}>1$} \\
					L & \mbox{if $i= \ID_{t+1}$ and $R^{(i)}_t=1$; $L$ obeys $\rho^{(i)}(L)$}
				\end{cases}, \\
				\eps^{(i)}_{t+1} &= 
					\begin{cases}
					\eps^{(i)}_t & \mbox{if $i \neq \ID_{t+1}$ or $R^{(i)}_{t}>1$} \\
						+1 & \mbox{with prob. $1/2$, if $i = \ID_{t+1}$ and $R^{(i)}_{t}=1$} \\
						-1 & \mbox{with prob. $1/2$, if $i = \ID_{t+1}$ and $R^{(i)}_{t}=1$}
				\end{cases}, \\ 
				\eps_{t+1} &= \eps_{t}^{(\ID_{t+1})}.
			\end{align}
		\end{subequations}
		Here the metaorder length is replenished according to the PDF $\{\rho^{(i)}(L)\}_L$ when the previous metaorder is terminated (i.e., if $R_t^{(\ID_{t+1})}=1$). 

	\subsection{Relationship with the original LMF model}
		Our model is a natural generalisation of the original LMF model to include the heterogeneity of the order-splitting behaviour. Indeed, our model reduces to the original LMF model by setting the parameter $\mathcal{P}$ as 
		\begin{equation}
			\lambda^{(i)} = \frac{1}{M}, \>\>\> \rho^{(i)}(L) = \rho(L) \>\>\> \mbox{for all } i \in \Omega_{\TR}
		\end{equation}
		by removing the heterogeneity in the order-splitting strategies. 

\section{Exact solutions}\label{sec:exactsol}
	In this section, we derive the exact solutions to our generalised LMF model. Particularly, we are interested in the order-sign autocorrelation function (ACF) in the stationary state: 
	\begin{equation}
		C_\tau := \lim_{t\to \infty}\la \eps_t\eps_{t+\tau}\ra = \la \eps_1\eps_{\tau+1}\ra_{\rm st}.
	\end{equation}

	\subsection{Preliminary calculation}
		\begin{figure}
			\centering
			\includegraphics[width=150mm]{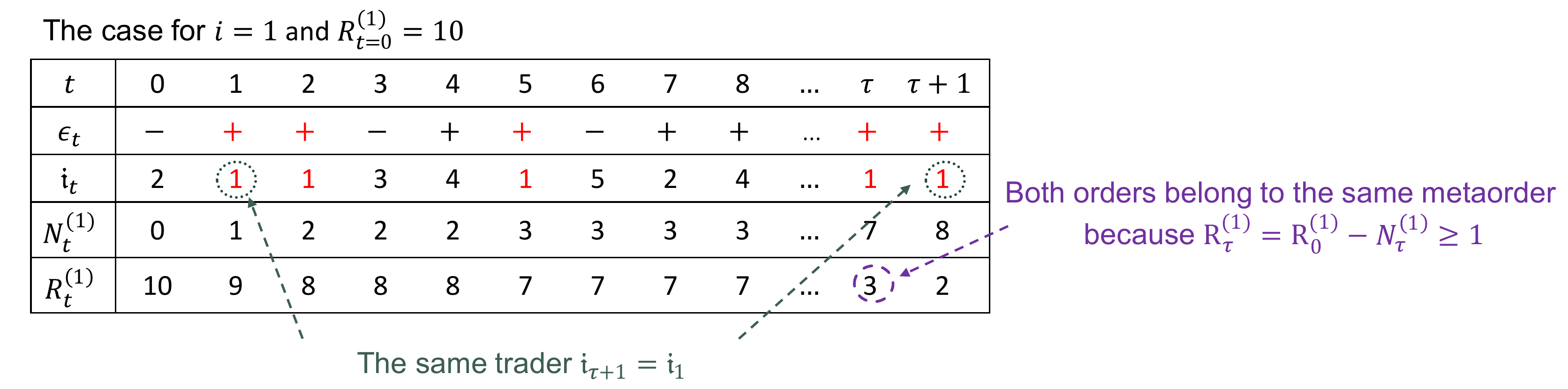}
			\caption{
				Schematic of the ACF decomposition for the case with $i=1$ and $R_{t=0}^{(i=1)}=10$. The issuer of the market orders at $t=1$ and $t=\tau+1$ is the same, such that $\ID_1=\ID_{\tau+1}=i=1$. Since $R_{t=\tau}^{(i=1)}=R_{t=0}^{(i=1)}-N_{t=\tau}^{(i=1)}=3\geq 1$, both orders at $t=1$ and $t=\tau+1$ belong to the same metaorder, and, thus, the condition of $u=1$ is met. 
			}
			\label{fig:R_0}
		\end{figure}
		%OK
		Before deriving the explicit formula of the exact ACF, we make a transformation of the definition of the ACF. Let us introduce a flag variable $u$ satisfying $u=1$ if the  metaorder executed at time $t=\tau+1$ belongs to the same metaorder executed at time $t=1$ or otherwise $u=0$. Let us introduce the conditioning on $u$, $\ID_{\tau+1}$, and $\ID_1$, to decompose the ACF as 
		\begin{equation}
			C_\tau = \sum_{u'\in \{0,1\}}\sum_{i\in \Omega_{\TR}}\sum_{j\in \Omega_{\TR}}\la \eps_1\eps_{\tau+1} | u=u', \ID_{\tau+1}=i, \ID_1=j \ra_{\rm st}P(u=u', \ID_{\tau+1}=i, \ID_1=j).
		\end{equation}
		See Fig.~\ref{fig:R_0} for a schematic of this decomposition.
		By construction, there is no correlation between the order signs belonging to two different metaorder. On the other hand, the order signs between the same metaorder are perfectly correlated. We thus obtain 
		\begin{equation}
			\la \eps_1\eps_{\tau+1} | u=u', \ID_{\tau+1}=i, \ID_1=j \ra_{\rm st}=\delta_{u',1}.
		\end{equation}
		In addition, $\ID_{\tau+1}$, and $\ID_1$ are independently generated, and 
		\begin{align}
			P(u=1, \ID_{\tau+1}=i, \ID_1=j) &= P(u=1 | \ID_{\tau+1}=i, \ID_1=j)P(\ID_{\tau+1}=i)P(\ID_1=j) \notag \\
			&= P(u=1 | \ID_{\tau+1}=\ID_1=i)\left(\lambda^{(i)}\right)^2\delta_{i,j}.
		\end{align}
		We obtain 
		\begin{equation}
			C_\tau = \sum_{i\in \Omega_{\TR}} \left(\lambda^{(i)}\right)^2P(u=1 | \ID_{\tau+1} = \ID_1=i).
		\end{equation}
		We next introduce the conditioning on $R_{t=0}^{(i)}$ as 
		\begin{equation}
			P(u=1 | \ID_{\tau+1} = \ID_1=i) = \sum_{R^{(i)}_0=2}^\infty P(u=1 | \ID_{\tau+1} = \ID_1=i, R_0^{(i)})P_{\rm st}(R_0^{(i)}),
		\end{equation}
		where we use the identity\footnote{
			See the following derivation: 
			\begin{equation}
				P(A|B) = \frac{P(A,B)}{P(B)} = \sum_{C} \frac{P(A,B,C)}{P(B)} = \sum_{C} \frac{P(A,B,C)}{P(B,C)}\frac{P(B,C)}{P(B)} = \sum_{C}P(A|B,C)P(C|B). 
				\label{eq:conditional_trans_1}
			\end{equation}
		}
		$P(A|B)=\sum_C P(A|B,C)P(C|B)$, and the relationships $P_{\rm st}(R_0^{(i)} | \ID_{\tau+1}=\ID_1=i)=P_{\rm st}(R_0^{(i)})$, and $P(u=1 | \ID_{\tau+1} = \ID_1=i, R_0^{(i)}=1)=0$.

		The term $P(u=1 | \ID_{\tau+1} = \ID_1=i, R_0^{(i)})$ is directly related with the survival probability of a metaorder whose initial volumes is $R_0^{(i)}$. Indeed, by defining $N^{(i)}_{\tau}$ as the total number of the metaorder executions by the trader $i$ during $[1,\tau]$, the condition of $u=1$ is equal to $R_0^{(i)}-N_{\tau}^{(i)}\geq 1$ (see Fig.~\ref{fig:R_0}), or equivalently, 
		\begin{equation}
			P(u=1 | \ID_{\tau+1} = \ID_1=i, R_0^{(i)}) = P(N^{(i)}_{\tau} \leq R_0^{(i)}-1).
		\end{equation}
		In other words, this is the probability that the discrete-time Poisson counting process $N^{(i)}_{\tau}$ remains within the range $[1,R_0^{(i)}-1]$ during the time interval $[1,\tau]$ with the initial condition $N_{t=1}^{(i)}=1$. 

		In summary, we can exactly decompose the total ACF as 
		\begin{screen}
		\begin{equation}
			C_\tau = \sum_{i\in \Omega_{\TR}} C_\tau^{(i)}, \>\>\> 
			C_\tau^{(i)}:= \left(\lambda^{(i)}\right)^2\sum_{R^{(i)}_0=2}^\infty P(N^{(i)}_{\tau} \leq R_0^{(i)}-1)P_{\rm st}(R_0^{(i)}),
			\label{eq:ACF_formula_basic}
		\end{equation}
		\end{screen}
		which needs the explicit formulas for $P_{\rm st}(R_0^{(i)})$ and $P(N^{(i)}_{\tau} \leq R_0^{(i)}-1)$. In the following subsections, we will derive the exact formulas for these quantities using the master equation approach. 

	\subsection{Stationary PDF for the remaining metaorder length}
		Let us derive the stationary PDF $P_{\rm st}(R^{(i)})$ for the remaining metaorder length $R^{(i)}$ via the master equation approach. 
		The master equation for the remaining metaorder length PDF $P_t(R^{(i)})$ is given by 
		\begin{align}
			\Delta_{t}P_t(R^{(i)}) = \lambda^{(i)}\left\{ P_t(R^{(i)}+1)-P_t(R^{(i)}) + P_t(1)\rho^{(i)}(R^{(i)})  \right\},\label{eq:FP_RDynamics}
		\end{align}
		where $\Delta_{t}P_t(R^{(i)}) := P_{t+1}(R^{(i)})-P_t(R^{(i)})$ for $R^{(i)}>0$ (see Appendix.~\ref{app:sec:der_ME_remainingMetaorderLength} for the derivation). In the stationary state $\Delta_{t}P_t\left(R^{(i)}\right) = 0$, we obtain the stationary distribution $P_{\rm st}(R^{(i)})$ in an exact form as 
		\begin{screen}
		\begin{align}
			P_{\rm st}(R^{(i)})&= c_R^{(i)} \rho_{\geq}^{(i)}(R^{(i)})\label{eq:StationaryR}
		\end{align}
		\end{screen}
		with the CCDF of $\rho^{(i)}_{\geq}(L)$ and the normalisation coefficient $c_R$ defined by 
		\begin{equation}
			\rho^{(i)}_{\geq}(L) := 1 - \sum^{L-1}_{L'=1}\rho^{(i)}(L') = \sum_{L'=L}^\infty \rho^{(i)}(L'), \>\>\>
			c_R^{(i)} = P_{\rm st}(1):=\frac{1}{\sum_{L=1}^\infty \rho_{\geq}^{(i)}(L)}.
		\end{equation}
		We note that $c_R$ can be transformed as 
		\begin{equation}
			\frac{1}{c_R^{(i)}} = \sum_{L=1}^\infty \rho_{\geq}^{(i)}(L) = \sum_{L=1}^\infty \sum_{L'=L}^\infty \rho^{(i)}(L') = \sum_{L'=1}^\infty \sum_{L=1}^{L'}\rho^{(i)}(L') =  \sum_{L'=1}^\infty L'\rho^{(i)}(L') = L_{\rm avg}^{(i)}.
		\end{equation}

	\subsection{Survival probability of the metaorder length}
		We next study the survival probability of the metaorder length, which is formulated as the CDF $P(N_{\tau}^{(i)}\leq R_0^{(i)}-1)$ of the discrete-time Poisson counting process $N_{t}^{(i)}$ with intensity $\lambda^{(i)}$ and initial condition $N_{t=1}^{(i)}=1$. Because the discrete-time Poisson counting process is equivalent to the Bernoulli process with probability $\lambda^{(i)}$, the PDF for $N_{t}^{(i)}$ obeys the binomial distribution\footnote{This solution can be systematically derived from the master-equation approach; see also Appendix~\ref{app:sec:ME_PoissonCountingProcess}.}: 
		\begin{equation}
			P_t(N^{(i)}) = \begin{cases}
				\mathcal{B}_{t-1,\lambda^{(i)}}(N^{(i)}-1) & (\mbox{for }N^{(i)} \in [1, t]) \\
				0 & (\mbox{for }N^{(i)} \not \in [1, t])
			\end{cases}
			\label{eq:P_t(N)_exactsol}
		\end{equation}
		with 
		\begin{equation}
			\mathcal{B}_{t,\lambda}(x) := \frac{t!}{x!(t-x)!}
				\lambda^{x}
				\left(1-\lambda\right)^{t-x},
		\end{equation}
		which satisfies the initial condition $P_{t=1}(N^{(i)}) = \delta_{N^{(i)},1}$. Finally, its CDF is given by 
		\begin{screen}
		\begin{equation}
			P(N_{\tau}^{(i)}\leq R_0^{(i)}-1) = \sum_{N^{(i)}=1}^{R_0^{(i)}-1} P_{\tau}(N^{(i)}) = \begin{cases}
				I_{1-\lambda^{(i)}}(\tau-R_{0}^{(i)}+1, R_{0}^{(i)}-1) & (\mbox{for }\tau > R_0^{(i)}-1) \\
				1 & (\mbox{for }\tau \leq R_0^{(i)}-1) 
			\end{cases}
		\end{equation}
		\end{screen}
		with the regularised incomplete Beta function $I_{x}(a,b)$ defined by 
		\begin{equation}
			I_{x}(a,b) := \frac{B(x;a,b)}{B(a,b)}, \>\>\> 
			B(x;a,b):= \int_0^x t^{a-1}(1-t)^{b-1}dt, \>\>\> 
			B(a,b) = B(1;a,b)
		\end{equation}
		for real numbers $x$, $a$, and $b$. 

	\subsection{Exact form of the order-sign ACF}
		We finally obtain the exact order-sign ACF formula in an explict form as 
		\begin{screen}
			\begin{subequations}
				\label{eq:exact_sol_ACF}
			\begin{align}
				C_\tau &= \sum_{i\in \Omega_{\TR}} C_\tau^{(i)}, \\
				C_\tau^{(i)}&= c_R^{(i)}\left(\lambda^{(i)}\right)^2\left[\sum_{R^{(i)}_0=2}^{\tau} \sum_{N^{(i)}=1}^{R_0^{(i)}-1} \rho_{\geq}^{(i)}(R^{(i)})\mathcal{B}_{t-1,\lambda^{(i)}}(N^{(i)}-1)+ \sum_{R^{(i)}_0=\tau+1}^\infty \rho_{\geq}^{(i)}(R^{(i)})\right] \notag \\
				&= c_R^{(i)}\left(\lambda^{(i)}\right)^2\left[\sum_{R^{(i)}_0=2}^{\tau} \rho_{\geq}^{(i)}(R^{(i)})I_{1-\lambda^{(i)}}(\tau-R_{0}^{(i)}+1, R_{0}^{(i)}-1) + \sum_{R^{(i)}_0=\tau+1}^\infty \rho_{\geq}^{(i)}(R^{(i)})\right].
			\end{align}
			\end{subequations}
		\end{screen}

	\subsection{Remark on the original derivation}
		Let us focus on the homogeneous case $\lambda^{(i)}=\lambda=1/M$ for all $i\in \Omega_{\TR}$. In the original LMF argument, they heuristically estimated that the original metaorder length at $\tau=1$ should obey the PDF 
		\begin{equation}
			Q(L) = \frac{L\rho(L)}{\sum_{L=1}^\infty L\rho(L)}
		\end{equation} 
		because a longer metaorder is likely to be observed with a higher probability. Furthermore, they assumed that the remaining metaorder length $R_{\tau=1}^{(i)}$ is uniformly distributed within $[1,L]$. On these heuristic but reasonable assumptions, they estimated the order-sign ACF as 
		\begin{equation}
			C_{\tau}^{\rm LMF} = \frac{1}{L_{\rm avg}} \sum_{L=1}^\infty\sum_{j=1}^{L-2}\sum_{h=0}^j \rho(L) \frac{(\tau-1)!}{h!(\tau-1-h)!}\lambda^{h+1}(1-\lambda)^{\tau-1-h}, \>\>\> L_{\rm avg}:= \sum_{L=1}^\infty L\rho(L).
		\end{equation}
		Our derivation is essentially similar to the original LMF argument. However, it is more systematic and rigorous version based on the master-equation approach without heuristic arguments. 
		
		Indeed, their heuristic formula is equivalent to ours except for a minor typo as follows: By exchange the sums between $L$ and $j$, we obtain 
		\begin{align}
			C_{\tau}^{\rm LMF} &= \frac{1}{L_{\rm avg}} \sum_{j=1}^{\infty}\sum_{h=0}^j\sum_{L=j+2}^\infty \rho(L) \frac{(\tau-1)!}{h!(\tau-1-h)!}\lambda^{h+1}(1-\lambda)^{\tau-1-h} \notag \\
			&= \frac{1}{L_{\rm avg}} \sum_{j=1}^{\infty}\sum_{h=0}^j \rho_{\geq}(j+2) \frac{(\tau-1)!}{h!(\tau-1-h)!}\lambda^{h+1}(1-\lambda)^{\tau-1-h} \notag \\
			&= \frac{1}{L_{\rm avg}} \sum_{R_0=3}^{\infty}\sum_{N=1}^{R_0-1} \rho_{\geq}(R_0) \frac{(\tau-1)!}{(N-1)!(\tau-N)!}\lambda^{N}(1-\lambda)^{\tau-N} \notag \\
			&= \frac{\lambda}{L_{\rm avg}} \sum_{R_0=3}^{\infty}\rho_{\geq}(R_0) P(N_{\tau}\leq R_0-1)
		\end{align}
		with formal replacements of the dummy variables between the second and third lines as $j=R_0-2$ and $h=N-1$. 
		
		By the way, for the homogeneous case, our exact formula~\eqref{eq:ACF_formula_basic} reduces to 
		\begin{equation}
			C^{\rm SK}_{\tau} = \frac{\lambda}{L_{\rm avg}}\sum_{R_0=2}^\infty \rho_{\geq}(R_0)P(N_{\tau}\leq R_0-1).
		\end{equation}
		Therefore, the LMF estimation $C^{\rm LMF}_{\tau}$ in Ref.~\cite{LMF_PRE2005} is consistent with our exact formula $C^{\rm SK}_{\tau}$ for the homogeneous case except for a very minor contribution from $R_0=2$. We think this minor contribution is just a typo without significant meanings, and our formula is a natural and rigorous extension of the original LMF theory.

\section{Examples and numerical verification}\label{sec:example}
		Let us derive the asymptotic behaviour of the order-sign ACF for several important cases. 

		\subsection{Case 0: Random traders}
			Let us consider the most trivial case where the trader $i$ submit her orders at purely random. This case corresponds to the setting 
			\begin{equation}
				\rho^{(i)}(L) = \delta_{L,1} \>\>\> \Longleftrightarrow \>\>\> 
				\rho_{\geq}^{(i)}(L) = \delta_{L,1} \>\>\> \mbox{for all } L\geq 1.
			\end{equation}
			From Eq.~\eqref{eq:exact_sol_ACF}, we obtain the order-sign ACF without any correlation as 
			\begin{equation}
				C_\tau^{(i)} = \delta_{\tau,1}.
			\end{equation}

		\subsection{Case 1: Exponential metaorder length distribution}
			\begin{figure}
				\centering
				\includegraphics[width=150mm]{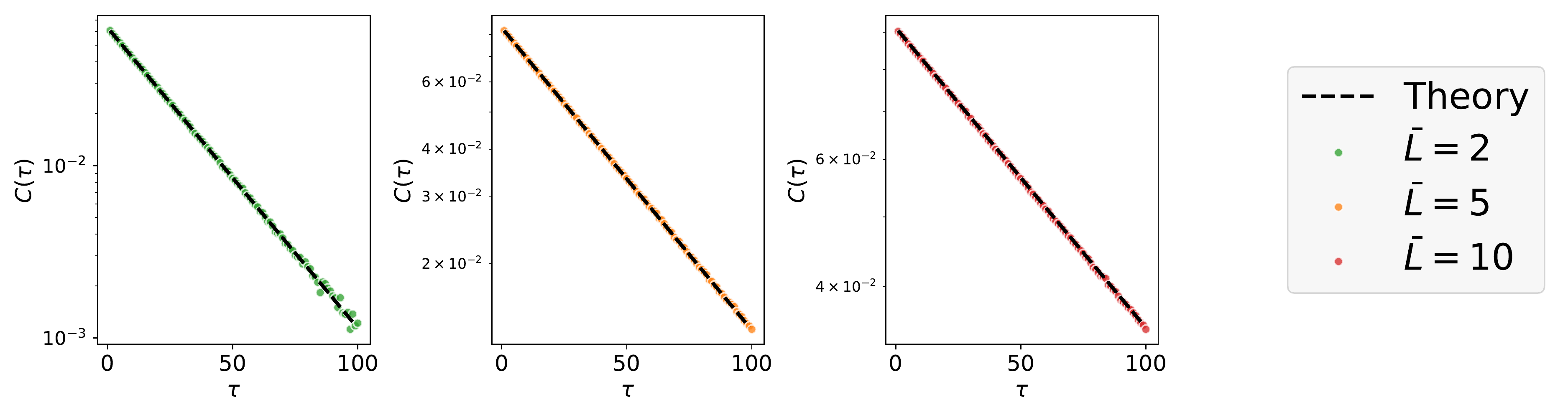}
				\caption{
					Comparisons between the numerical results and the theoretical prediction~\eqref{eq:ACF_exact_expon} by assuming that all the traders are the exponentially-splitting traders with the same parameters $L^{*(i)}=\bar{L}$, $\lambda^{(i)}=\bar{\lambda}$ for all $i\in \Omega_{\rm TR}$. The market ACF is theoretically given by $C_\tau = \sum_{i\in \Omega_{\rm TR}}C_\tau^{(i)}=\bar{\lambda}e^{-1/\bar{L}}(1-\bar{\lambda}(1-e^{-1/\bar{L}}))^{\tau-1}$ for $\tau \geq 1$. The numerical autocorrelation functions of the generalised LMF model are shown for the case (a) with $(M,\bar{L},\bar{\lambda})=(10,2,0.1)$ as the green line, the case (b) with $(M,\bar{L},\bar{\lambda})=(10,5,0.1)$ as orange line, and the case (c) with $(M,\bar{L},\bar{\lambda})=(10,10,0.1)$ as the red line. 
				}\label{fig:LMFExpon}
			\end{figure}
			Let us consider the case where the metaorder length obeys the exponential law: 
			\begin{equation}
				\rho_{\geq}^{(i)}(L) = e^{-(L-1)/L^{*(i)}}, \>\>\> L^*>0 \>\>\> \Longleftrightarrow \>\>\> 
				\rho^{(i)}(L) = \left(e^{1/L^{*(i)}}-1\right)e^{-L/L^{*(i)}}.
				\label{eq:expon_traders_assump}
			\end{equation} 
			Note that $c_{R}^{(i)} := 1/\sum_{L=1}^\infty \rho_{\geq}^{(i)}(L) = 1-e^{-1/L^{*(i)}}$. For this case, we obtain an exact ACF formula, such that 
			\begin{equation}
				C_\tau^{(i)} = \left(\lambda^{(i)}\right)^2e^{-1/L^{*(i)}}\left(1-\lambda^{(i)}+\lambda^{(i)}e^{-1/L^{*(i)}}\right)^{\tau-1} \>\>\>\mbox{for }\tau\geq 1.
			\end{equation}
			See Appendix~\ref{sec:app:ACF_exact_expon} for the detailed derivation. 
			This equation can be rewritten as 
			\begin{screen}
			\begin{subequations}
				\label{eq:ACF_exact_expon} 
				\begin{equation}
					C_\tau^{(i)} = c_{\rm ET}^{(i)}e^{-\tau/\tau^{*(i)}}\>\>\mbox{for $\tau\geq 1$}	
				\end{equation}
				with 
				\begin{equation}
					\>\>
					c_{\rm ET}^{(i)} := \frac{\left(\lambda^{(i)}\right)^2e^{-1/L^{*(i)}}}{1-\lambda^{(i)}\left(1-e^{-1/L^{*(i)}}\right)}, \>\>\>
					\frac{1}{\tau^{*(i)}} := \ln \frac{1}{1-\lambda^{(i)}\left(1-e^{-1/L^{*(i)}}\right)}.
				\end{equation}
			\end{subequations}
			\end{screen}
			This implies that the exponential decay appears in the order-sign ACF as a fast-decaying tail, which is consistent with empirical observations. We numerically checked the validity of this formula as shown in Fig.~\ref{fig:LMFExpon}. 

		\subsection{Case 2: Power-law metaorder length distribution}
			\begin{figure}
				\centering
				\includegraphics[width=150mm]{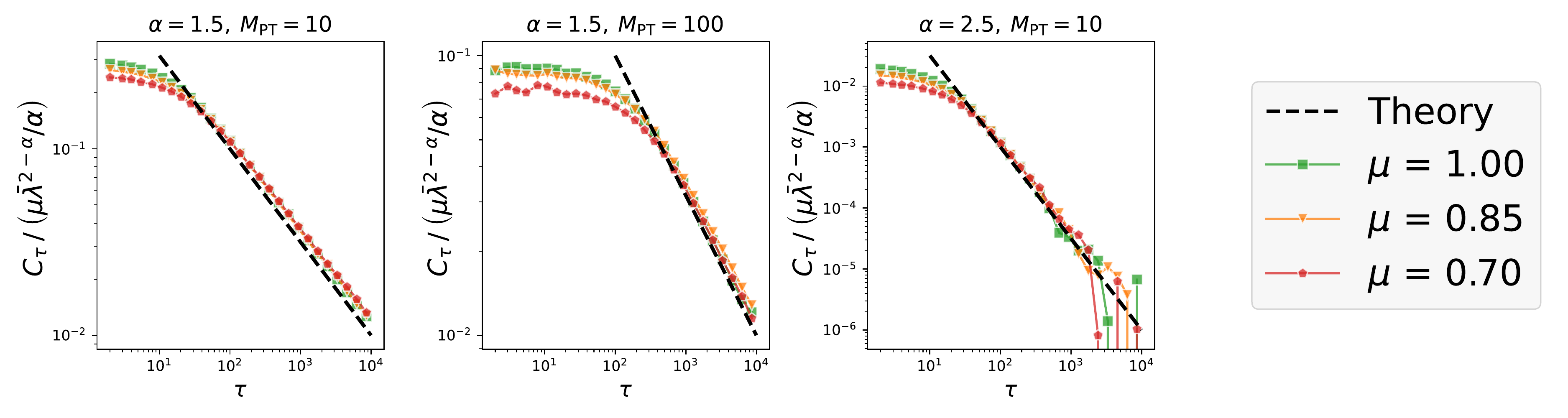}
				\caption{
					Comparisons between the numerical results and the theoretical prediction when both random and power-law splitting traders coexist with $|\Omega_{\rm PL}|=M_{\rm PT}$, $\alpha^{(i)}=\alpha$, and $\lambda^{(i)}=\bar{\lambda}:=\mu/M_{\rm PT}$ for all $i\in \Omega_{\rm PT}$. The total intensity of the random-trader submission is $1-\mu$. The green, orange, and red lines shows the results for $\mu=1.00$, $\mu=0.85$, and $\mu=0.70$, respectively. The market ACF is theoretically given by $C_\tau = \sum_{i\in \Omega_{\rm TR}}C_\tau^{(i)}\simeq (\mu\bar{\lambda}^{2-\alpha}/\alpha)\tau^{-\alpha+1}$ for $\tau \gg 1$. We used the following parameters: $(\alpha, M)=(1.5,10)$ for Fig.~(a), $(\alpha,M_{\rm PT})=(1.5,100)$ for Fig.~(b), and $(\alpha,M_{\rm PT})=(2.5,10)$ for Fig.~(c). 
				}
				\label{fig:LMFCoexist}
			\end{figure}
			We next study the case where the metaorder length obeys the power law: 
			\begin{equation}
				\rho^{(i)}_{\geq}(L) = L^{-\alpha^{(i)}}
				\label{eq:power-law_length_assump}
			\end{equation}
			with a positive constant $\alpha^{(i)}>1$. This means that the density profile is approximately given by 
			\begin{equation}
				\rho^{(i)}(L) \simeq -\frac{\rho^{(i)}_{\geq}(L)}{dL} = \alpha^{(i)} L^{-\alpha^{(i)}-1}.
			\end{equation}
			For this case, by using an integral approximation of the sum, we obtain 
			\begin{equation}
				\frac{1}{c_R^{(i)}} = L_{\rm avg} := \sum_{L=1}^\infty L \rho(L) \simeq \int_1^\infty \alpha^{(i)}L^{-\alpha^{(i)}}dL = \frac{\alpha^{(i)}}{\alpha^{(i)}-1}.
			\end{equation}
			For sufficiently large $\tau \gg 1$, we asymptotically obtain 
			\begin{screen}
				\begin{equation}
					C_{\tau}^{(i)} \simeq c_{\rm PT}^{(i)}\tau^{-\gamma^{(i)}}, \>\>\> \gamma^{(i)} := \alpha^{(i)}-1, \>\>\> c_{\rm PT}^{(i)}:=\frac{\left(\lambda^{(i)}\right)^{3-\alpha^{(i)}}}{\alpha^{(i)}}.
					\label{eq:LMF_powerlaw_multi_gen}
				\end{equation}
			\end{screen}
			For the detailed derivation, see Appendix~\ref{sec:app:ACF_asymp_powerlaw}. 

	\subsection{ACF formula with heterogeneous strategies}
		Let us summarise the above formula regarding the heterogeneity of the order-splitting strategies. Let us consider a market where the following-types of traders coexist: 
		\begin{itemize}
			\item random traders (whose set is denoted by $\Omega_{\rm RT}$), 
			\item exponentially-splitting traders (whose set is denoted by $\Omega_{\rm ET}$), and 
			\item power-law splitting traders (whose set is denoted by $\Omega_{\rm PT}$). 
		\end{itemize}
		The total ACF asymptotically obeys 
		\begin{screen}
			\begin{equation}
				C_\tau \simeq \sum_{i\in \Omega_{\rm ET}}c_{\rm ET}^{(i)}e^{-\tau/\tau^{*(i)}} + \sum_{i\in \Omega_{\rm PT}}c_{\rm PT}^{(i)}\tau^{-\gamma^{(i)}}.
				\label{eq:hetero_ACF_formula}
			\end{equation}
		\end{screen}
		Thus, while we observe the fast decay characterised by the exponential law for relatively small $\tau$, the slow decay is dominant for large $\tau$. Such characters are consistent with the empirical observations. 
		
		\subsubsection{Remark 1: consistency with the original LMF formula for the homogeneous case}
			Let us assume that all traders are power-law splitting traders with homogeneous intensity, such that 
			\begin{equation}
				\alpha^{(i)} = \alpha, \>\>\> \lambda^{(i)} = \frac{1}{M} \>\>\> \mbox{for all }i\in \Omega_{\rm PT} = \Omega_{\rm TR}.
			\end{equation}
			For this case, we obtain 
			\begin{equation}
				C_\tau \simeq \frac{\tau^{-\gamma}}{\alpha M^{2-\alpha}}, \>\>\> \gamma := \alpha-1, 
			\end{equation}
			which is equivalent to the original LMF formula~\eqref{eq:LMF_original_LRC}.

		\subsubsection{Remark 2: the importance of the minimum power-law exponent $\alpha_{\min}$}
			The ACF is finally characterised by the power law 
			\begin{equation}
				C_\tau \propto \tau^{-\alpha_{\min}+1}\>\>\> 
				\mbox{for large }\tau, \>\>\> 
				\alpha_{\min} := \min_{i\in \Omega^{(i)}} \alpha^{(i)}
			\end{equation}
			Thus, $\alpha_{\min}$ is the most important parameter characterising the final asymptotic behaviour of the ACF. 

			This character is relevant to the data calibration. Indeed, a typical quantity that is empirically-available is the aggregated metaorder-length distribution among all the splitting traders $\Omega_{\rm ST}:=\Omega_{\rm ET} + \Omega_{\rm PT}$, such that 
			\begin{equation}
				\rho_{\rm ST}^{\rm empirical}(L) := \frac{1}{N_{\rm tot}}\sum_{k=1}\delta(L-L_k),
				\label{def:empirical_L_aggregated_PDF}
			\end{equation}
			where $N_{\rm tot}$ is the total number of metaorder lengths and $L_k$ is the $k$th metaorder length among all the splitting traders. Let us decompose this aggregated empirical distribution as 
			\begin{align}
				\rho_{\rm ST}^{\rm empirical}(L) = \frac{1}{N_{\rm tot}}\sum_{i\in \Omega_{\rm ST}}\sum_{k=1}^{N_{\rm tot}^{(i)}}\delta(L-L_k^{(i)}) = \sum_{i\in \Omega_{\rm ST}}\frac{N_{\rm tot}^{(i)}}{N_{\rm tot}}\left\{\frac{1}{N_{\rm tot}^{(i)}}\sum_{k=1}^{N_{\rm tot}^{(i)}}\delta(L-L_k^{(i)})\right\}
			\end{align}
			with $L_{k}^{(i)}$ being the $k$th metaorder length of the trader $i$ and $N_{\rm tot}^{(i)}$ being the total number of metaorder lengths of the trader $i$. 
			Here we use the ergodicity regarding the empirical distributions
			\begin{equation}
				\rho^{(i)}(L) = \lim_{N_{\rm tot}^{(i)} \to \infty}\frac{1}{N_{\rm tot}^{(i)}}\sum_{k=1}^{N_{\rm tot}^{(i)}}\delta(L-L_k^{(i)}).
			\end{equation}
			Also, we can evaluate the following quantities for a long-time simulation with the simulation time $t$ as 
			\begin{equation}
				N_{\rm tot}^{(i)} \simeq \frac{\lambda^{(i)}t}{\la L_i\ra}, \>\>\> 
				N_{\rm tot} \simeq \frac{\lambda_{\rm ST} t}{\la L\ra}, \>\>\> 
				\lambda_{\rm ST} := \sum_{i\in \Omega_{\rm ST}}\lambda^{(i)}.
			\end{equation}
			We thus obtain 
			\begin{equation}
				\rho_{\rm ST}^{\rm empirical}(L) \simeq \sum_{i\in \Omega_{\rm ST}} w^{(i)}\rho^{(i)}(L) \propto L^{-\alpha^{\min}-1} \>\> \mbox{for a large $L$ with} \>\> w^{(i)} := \frac{\lambda^{(i)}}{\lambda_{\rm ST}}\frac{\la L\ra}{\la L_i\ra}.\label{eq:MarketAlpha}
			\end{equation}
			This relation implies that it is acceptable to use the aggregated metaorder-length distributions among all the splitting traders in determining $\alpha_{\min}$. 

			We note that these formulas are derived by assuming that the sample size is infinity and the metaorder-length distributions obey the true power-law without cutoffs. Technically, the straightforward applicability of these formulas is rather limited for real data analyses, where the sample size is finite and the metaorder-length PDFs obey truncated power laws. However, the above analysis highlights the conceptual relation between the metaorder-length PDF for individual traders and the aggregated PDF that is empirically accessible.

\section{Theoretical discussion 1: data calibration based on the power-law splitting assumption}\label{sec:discussion}
%\color{black}

	Here we discuss the implication of our heterogeneous-LMF formula~\eqref{eq:hetero_ACF_formula} for the data calibration. Particularly, in this section, we only make a simple assumption
	\begin{equation}
		\alpha^{(i)} = \alpha \>\> \mbox{for all }i \in \Omega_{\rm PT}
	\end{equation}
	with the heterogeneity included in the intensities $\{\lambda^{(i)}\}_{i \in \Omega_{\rm PT}}$ among the power-law splitting traders. Also, the total intensity $\mu$ and the total number $M$ of the power-law splitting traders are denoted by 
	\begin{equation}
		\mu := \sum_{\i\in \Omega_{\rm PT}} \lambda^{(i)}, \>\>\> 
		M_{\rm PT} := |\Omega_{\rm PT}|,
	\end{equation} 
	respectively. For this case, the asymptotic behaviour is described by 
	\begin{screen}
	\begin{equation}
		C_\tau \simeq c_0^{\rm SK}\tau^{-\gamma}, \>\>\> 
		\gamma := \alpha-1, \>\>\> 
		c_0^{\rm SK} = \frac{1}{\alpha}\sum_{i\in \Omega_{\rm PT}}\left(\lambda^{(i)}\right)^{3-\alpha}.
		\label{eq:SK_hetero_discussion}
	\end{equation}
	\end{screen}

	\subsection{Robust power-law exponent formula}
		What is the implication of the heterogeneous LMF formula~\eqref{eq:SK_hetero_discussion} for the data calibration? One of the most important implications is that the power-law exponent $\gamma$ is insensitive to the heterogeneity of the intensity distribution $\{\lambda^{(i)}\}_{i \in \Omega_{\rm PT}}$. This is a very important character of our heterogenous LMF model because it implies that the power-law exponent is a very robust measurable quantity: even if the heterogeneity of the intensity distribution is present, the LMF prediction $\gamma = \alpha-1$ is a trustable relationship. In real datasets, the average waiting time $\tau^{(i)}:= 1/\lambda^{(i)}$ is expected to distribute widely, such as the power-law distribution $P(\tau):=(1/M)\sum_{i\in \Omega_{\rm PT}}\delta(\tau-\tau^{(i)})\propto \tau^{-\chi-1}$ for large $\tau$ with $\chi > 0$. This assumption is equivalent to the power-law peak asymptotics in the intensity distribution, such that $P(\lambda)\propto \lambda^{\chi-1}$ for small $\lambda$. The relation~\eqref{eq:SK_hetero_discussion} states that such widely-distributed waiting times (or intensities) have no impact on the macroscopic power-law exponent $\gamma$ in the ACF, which is non-trivial. From this viewpoint, the heterogeneity in the LMF model is not essential in understanding the LRC; the original LMF is a sufficient model. 

		In addition, since the formula $\gamma=\alpha-1$ does not depend on the intensities $\{\lambda^{(i)}\}_i$, the power-law ACF formula will hold even when the intensities have slow time dependence if $\alpha$ is time independent. Indeed, in the presence of the time inhomogeneity of $\{\lambda^{(i)}(t)\}_i$, the ACF formula will be replaced by 
		\begin{equation}
			C_{\tau} \simeq \bar{c}_0^{\rm SK} \tau^{-\gamma}, \>\>\> 
			\gamma=\alpha-1, \>\>\> 
			\bar{c}_0^{\rm SK} := \frac{1}{\alpha T_{\rm fin}}\sum_{i\in \Omega_{\rm PT}}\int_0^{T_{\rm fin}}  \left(\lambda^{(i)}(t)\right)^{3-\alpha}dt 
		\end{equation}
		with the final observation time $T_{\rm fin}$, if the time dependence of $\{\lambda^{(i)}(t)\}_i$ is sufficiently slow. Given that the intensities $\{\lambda^{(i)}(t)\}_i$ will change day by day, it is pleasant that the power-law ACF formula holds independently of the intensities, at least for their slow time variation.

	\subsection{Non-robust prefactor formula}
		On the other hand, the prefactor $c_0^{\rm SK}$ is very sensitive to the heterogeneity of the intensity distribution $\{\lambda^{(i)}\}_{i \in \Omega_{\rm PT}}$. Indeed, the prefactor $c_0^{\rm SK}$ is different from the homogeneous LMF formula: 
		\begin{equation}
			c_0^{\rm SK} \neq \frac{1}{\alpha M_{\rm PT}^{2-\alpha}},
		\end{equation}
		if $\lambda^{(i)}\neq 1/M_{\rm PT}$ for some $i \in \Omega_{\rm PT}$. This implies that the interpretation of the prefactor is not straightforward because it sensitively depends on the underlying microscopic assumptions. We are sure that the homogenous assumption in the intensity distribution is unrealistic in real datasets. 

		Furthermore, we have assumed that the metaorder length PDF exactly obeys the paretian distribution for all the range. This assumption is also unrealistic. Rather, it is a more realistic assumption that the power-law holds only asymptotically: 
		\begin{equation}
			\rho_{\geq}(L) \simeq c_{\rho} L^{-\alpha} \>\>\> \mbox{for large }L.
		\end{equation}
		Actually, we validated this weaker assumption in our microscopic datasets of the TSE market (see Ref.~\cite{SatoPRL2023,SatoPRR2023}). Under this assumption, the prefactor is slightly modified. Anyway, the prefactor is sensitive to the model-specific assumptions.

	\subsection{Systematic underestimation of the prefactor by the homogeneous LMF model}
		Furthermore, our heterogeneous LMF formula~\eqref{eq:SK_hetero_discussion} implies that the prefactor is systematically biased in the presence of the heterogeneous intensities. To clarify this point, let us consider the homogenous assumption in the intensities among the power-law splitting traders, such that 
		\begin{equation}
			\lambda^{(i)}_{\rm LMF} = \frac{\mu}{M_{\rm PT}},
		\end{equation}
		while we assume that the random and exponentially-splitting traders can be present (i.e., $\mu$ can be different from the unity). For this case, the prefactor is given by 
		\begin{equation}
			c_0^{\rm LMF} = \frac{\mu^{3-\alpha}}{\alpha M_{\rm PT}^{2-\alpha}},
		\end{equation}
		which reduces to Eq.~\eqref{eq:LMF_original_prefactor_LRC} for $\mu=1$. Here we can prove that the prefactor is systematically underestimated by the homogeneous LMF model with $\alpha \in (1,2)$: 
		\begin{screen}
			\begin{equation}
				c_0^{\rm LMF} \leq
				c_0^{\rm SK}  \leq \frac{\mu^{3-\alpha}}{\alpha} .
				\label{eq:inequality_SKvsLMF}
			\end{equation}
		\end{screen}
		The lower-bound equality holds when the intensities are homogeneous, such that $\lambda^{(i)}=\mu/M_{\rm PT}$ for all $i\in \Omega_{\rm PT}$. In addition, the upper-bound equality holds when the power-law splitter is alone $M_{\rm PT}=1$. 

		These inequalities highlight the impact of the heterogeneous trading strategies on the prefactor estimation, and is the final main result of this report. The inequality~\eqref{eq:inequality_SKvsLMF} is practically relevant to the evaluation of the ACF prefactors by data calibration. 

		\paragraph{Proof.}
			The lower bound in the inequality~\eqref{eq:inequality_SKvsLMF} is proved by the H\"older's inequality: 
			\begin{equation}
				\left(\sum_{i} |a_i|^{p}\right)^{1/p} \left(\sum_{i} |b_i|^{q}\right)^{1/q} \geq \sum_{i} |a_ib_i|  
			\end{equation}
			for any series $\{a_i\}_i, \{b_i\}_i$ and any real numbers $p, q$ satisfying $1/p+1/q=1$, $p \geq 1$, and $q \geq 1$. By putting $a_i = \lambda^{(i)}$, $b_i = \lambda^{(i)}_{\rm LMF}=\mu/M_{\rm PT}$, and $p = 3-\alpha \in (1,2)$, we obtain 
			\begin{equation}
				\left\{\sum_{i\in \Omega_{\rm PT}} \left(\lambda^{(i)}\right)^{p}\right\}^{1/p} \left\{\sum_{i\in \Omega_{\rm PT}} \left(\frac{\mu}{M_{\rm PT}}\right)^{q}\right\}^{1/q} \geq \sum_{i\in \Omega_{\rm PT}} \lambda^{(i)}\frac{\mu}{M_{\rm PT}},
			\end{equation}
			which is equivalent to 
			\begin{equation}
				\sum_{i\in \Omega_{\rm PT}} \left(\lambda^{(i)}\right)^{3-\alpha} \geq \frac{\mu^{3-\alpha}}{M_{\rm PT}^{2-\alpha}}. 
			\end{equation}
			We thus obtain the inequality~\eqref{eq:inequality_SKvsLMF} for the lower bound. 

			We next prove the upper bound in the inequality~\eqref{eq:inequality_SKvsLMF}. Given that $\lambda^{(i)} \geq 0$ and $3-\alpha \in (1,2)$, we apply the inequality~\eqref{app:ineq2} in Appendix~\ref{sec:app:inequality_2} by setting $x_i = \lambda^{(i)}$. We then obtain 
			\begin{equation}
				\mu^{3-\alpha} = \left(\sum_{i \in \Omega_{\rm PT}} \lambda^{(i)}\right)^{3-\alpha} \geq  
				\sum_{i \in \Omega_{\rm PT}} \left(\lambda^{(i)} \right)^{3-\alpha},
			\end{equation}
			which is equivalent to the upper-bound inequality for \eqref{eq:inequality_SKvsLMF}. 

	\subsubsection{Estimation of the lower bound of the total number of order-splitting traders}
		The inequality~\eqref{eq:inequality_SKvsLMF} is useful for the estimation of the total number of order-splitting traders. Indeed, we can estimate the lower bound of the total number of the power-law order-splitting traders as   
		\begin{screen}
			\begin{equation}
				M^{\rm LB}_{\rm PT} \lesssim M_{\rm PT}, \>\>\> 
				M^{\rm LB}_{\rm PT} := \left(\frac{\mu^{3-\alpha}}{\alpha c_0^{\rm dat}}\right)^{\frac{1}{2-\alpha}}
				\label{eq:inequality_M}
			\end{equation}
		\end{screen}
		by assuming $c_{0}^{\rm SK}\simeq c_0^{\rm dat}$, where $c_0^{\rm dat}$ is the empirically-available prefactor. Since $\gamma$ is directly measurable from the ACF, $\alpha$ is also indirectly measurable by the relationship $\alpha=\gamma+1$. While $\mu$ is not an empirically observable quantity from public data, Ref.~\cite{SatoPRL2023,SatoPRR2023} reports that $\mu$ was typically $0.8$ in the Tokyo Stock Exchange market from 2012 to 2020. Thus, it might be possible to roughly evaluate the lower bound of the total number of order-splitting traders $M_{\rm PT}$ from this inequality.

	%%%%--------------------------------------------------%%%%%
%\color{blue}
\section{Theoretical discussion 2: superposition of the exponential splitting traders}
\label{sec:discussion2}
	In Sec.~\ref{sec:discussion}, we discuss the theoretical scenario in the overwhelming presence of power-law splitting traders to understand the origin of the LRC in the market-order flow. While this scenario is the most promising and plausible, here we discuss other theoretical possibilities that the LRC appears as the superposition of exponential splitting traders. In other words, we assume the absence of powew-law splitting traders $M_{\rm PT}=0$ but the dominant presence of exponential splitting traders $M_{\rm ET}\neq 0$. Interestingly, the LMF prediction $\gamma = \alpha-1$ still holds even for this alternative scenario, suggesting the robustness of the LMF prediction. 
	
	Let us define the empirical distribution function of $(L^{*(i)}, \lambda^{(i)})$, which characterises the exponential splitting traders: 
	\begin{equation}
		P_{\rm ET}(L^*,\lambda) := \frac{1}{M_{\rm ET}}\sum_{i\in \Omega_{\rm ET}}\delta(L^*-L^{*(i)})\delta(\lambda-\lambda^{(i)}), \>\>\> M_{\rm ET} := |\Omega_{\rm ET}|.
		\label{eq:def_continuousP_ET}
	\end{equation} 
	For simplicity, we assume that $M_{\rm ET}$ is large enough for $P_{\rm ET}(L^*,\lambda)$ to be approximated as a continuous function and that $L^*$ and $\lambda$ are statistical independent, such that 
	\begin{equation}
		P_{\rm ET}(L^*,\lambda) = P_{\rm ET}(L^*)P_{\rm ET}(\lambda)
		\label{eq:approx_factorizeP_ET} 
	\end{equation}
	with $P_{\rm ET}(L^*):=(1/M_{\rm ET})\sum_{i\in \Omega_{\rm ET}} \delta(L^*-L^{*(i)})$ and $P_{\rm ET}(\lambda):=(1/M_{\rm ET})\sum_{i\in \Omega_{\rm ET}} \delta(\lambda-\lambda^{(i)})$. In addition, we assume the total number of exponential splitting traders is sufficiently large $M_{\rm ET}\gg 1$ and there is no single trader overwhelmingly contributing to the total market orders, such that 
	\begin{equation}
		\lambda^{(i)} \ll 1 \>\> \mbox{for all }i\in \Omega_{\rm ET}.
		\label{eq:approx_noOverwhelmingET} 
	\end{equation}

	\subsection{Scenario based on the fat-tailed decay length distribution}
		\begin{figure}
			\centering
			\includegraphics[width=150mm]{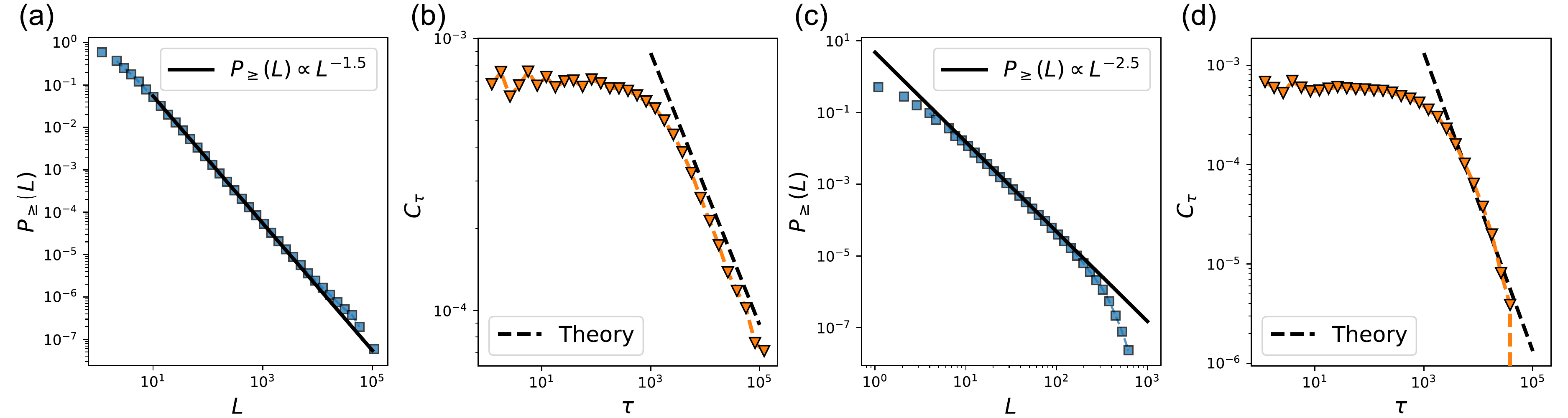}
			\caption{
				Comparisons between the numerical results and the theoretical prediction in the system constituted by exponential splitting traders with various decay lengths. 
				(a-b)~The aggregated metaorder length distribution and the autocorrelation function of the order flow under $|\Omega_{\rm EX}|=10^3$, $\lambda^{(i)}=|\Omega_{\rm EX}|^{-1}$, and $P(L^{*(i)})=\left(L^{*(i)}\right)^{-0.5}$.
				(c-d)~The aggregated metaorder length distribution and the autocorrelation function of the order flow under $|\Omega_{\rm EX}|=10^3$, $\lambda^{(i)}=|\Omega_{\rm EX}|^{-1}$, and $P(L^{*(i)})=\left(L^{*(i)}\right)^{-1.5}$. 
			}
			\label{fig:LMFSuperPosition-RunDist}
		\end{figure}

		Let us focus on the strong inhomogeneity in the decay length $L^{*(i)}$, such that 
		\begin{equation}
			P(L^*)\simeq (	\vartheta-1)\left(L^{*}\right)^{-\vartheta}
		\end{equation}
		with $\vartheta \in (1,2)$. On this assumption, we find that both the market-order ACF $C_{\tau}$ and the aggregated empirical metaorder-length PDF $\rho_{\rm ST}^{\rm empirical}(L)$, defined by Eq.~\eqref{def:empirical_L_aggregated_PDF}, obeys the power law (see Fig.~\ref{fig:LMFSuperPosition-RunDist}; see Appendices~\ref{sec:app:ACF-Ex-Superposition} and~\ref{sec:app:SimMethod} for the derivation and the technical details of numerical simulations): 
		\begin{screen}
			\begin{subequations}\label{eq:superposition-RunDist}
				\begin{equation}
					C_\tau \simeq  q_0^{\rm SK}\tau^{-\gamma},\>\>\> 
					\gamma = \alpha - 1, \>\>\> 
					q_0^{\rm SK} := \Gamma(\alpha)\sum_{i=1}^{M_{\rm ET}} \left(\lambda^{(i)}\right)^{3-\alpha},
					\label{eq:superposition-RunDist_1}
				\end{equation}
				\begin{equation}
					\rho_{\rm ET}^{\rm empirical}(L) \propto L^{-\alpha-1},\>\>\>\alpha = \vartheta.
					\label{eq:superposition-RunDist_2}
				\end{equation}
			\end{subequations}
		\end{screen}

		\subsubsection{Relationship to the previous BBDG model}
			Let us discuss the relationship to the previous model in the textbook~\cite{BouchaudB} by Bouchaud, Bonart, Donier, and Gould (BBDG). BBDG proposed a variant of the LMF model (which we call the BBDG model in this article to distinguish the two models) to simplify the algebraic calculations for the ACF formulas. The BBDG model is based on the assumption that the stopping intensity of order splitting obeys a power law. Our scenario based on superposition of exponential splitters is essentially similar to the BBDG model. Indeed, by setting $\lambda^{(i)}=\mu/M_{\rm ET}$, we obtain 
			\begin{equation}
				q_0^{\rm BBDG} := \Gamma(\alpha)\frac{\mu^{3-\alpha}}{M_{\rm ET}^{2-\alpha}},
			\end{equation}
			which is equivalent to the formula in Ref.~\cite{BouchaudB} when $\mu=1$. 

		\subsubsection{Robustness of the power-law formula}
			The results~\eqref{eq:superposition-RunDist} highlight the robustness of the LMF power-law prediction $\gamma=\alpha-1$: if the empirical aggregated metaorder-length PDF $\rho_{\rm ST}^{\rm empirical}(L)$ obeys the power law with exponent $\alpha$, we can expect the ACF power-law decay with exponent $\gamma=\alpha-1$ whether $\rho_{\rm ST}^{\rm empirical}(L)$ is composed of power-law splitters or superposition of exponential-law splitters. This prediction is robust and insensitive to the details of the underlying microscopic dynamics, even regarding the types of splitters (regardless of whether they are power-law or exponential-law splitters). This character is pleasant and reliable for data analyses.

		\subsubsection{Robustness of the prefactor formulas}
			The prefactor formula~\eqref{eq:superposition-RunDist} is very similar to the prefactor formula~\eqref{eq:SK_hetero_discussion} in Sec.~\ref{sec:discussion}. Indeed, we have 
			\begin{screen}
				\begin{equation}
					q_0^{\rm BBDG} \leq
					q_0^{\rm SK}  \leq \mu^{3-\alpha} \Gamma(\alpha) .
					\label{eq:inequality_SKvsBBDG}
				\end{equation}
			\end{screen}
			In other words, the prefactor is smallest if and only if the submission intensities are homogeneous, such that $\lambda_i=\mu/M_{\rm ET}$ for all $i\in \Omega_{\rm ET}$.

		\subsubsection{Remark on the essential similarity between the LMF and BBDG models}
			In addition, we find that the prefactor formulas between $c_0$ and $q_0$ are essentially similar in the sense that 
			\begin{screen}
				\begin{equation}
					1 \leq \frac{q_0^{\rm BBDG}}{c_0^{\rm LMF}} = \frac{q_0^{\rm SK}}{c_0^{\rm SK}} = \alpha \Gamma(\alpha) \leq 2 \>\>\> \mbox{for }\alpha \in (1,2).
				\end{equation}
			\end{screen}
			In other words, the prefactors are almost the same between the two scenarios except for factor 2 at most. 

	\subsection{Open question: the power-law splitter scenario vs. the superposed exponential-law splitter scenario}
		We presented various theoretical scenarios to derive the market ACF from the order-splitting hypothesis. For example, the power-law exponent formula $\gamma=\alpha-1$ and the prefactor inequality~\eqref{eq:inequality_SKvsLMF} robustly hold for both two scenarios, in the presence of power-law splitters or superposition of exponential splitters with various decay length. Note that the power-law ACF can be derived even for the scenario of superposition of exponential splitters with various intensities (see Appendix~\ref{sec:app:ACF-Ex-Superposition-Intensity}). 

		The natural question is which scenario is most plausible in reality, the power-law splitter scenario or the superposition of the exponential-law splitters. While the robustness of the LMF predictions is a pleasant character in applying the LMF theory, verifying the LMF predictions itself does not immediately imply the rejection of either scenario inversely due to the robustness of the LMF formulas. In other words, our previous reports~\cite{SatoPRL2023,SatoPRR2023} establish the relationship $\gamma=\alpha-1$ and the inequality~\eqref{eq:inequality_SKvsLMF}, but, technically, they do not distinguish the two scenarios of the power-law splitting and the superposition of exponential splitting. It is an important issue to reject either scenario by analysing financial microscopic datasets from different perspectives. We leave this problem as an open question for the next data-analytical step. 

\section{Conclusion}\label{sec:conclusion}
	We have proposed a generalised Lillo-Mike-Farmer model by incorporating the heterogeneity of order-splitting strategies. This model is exactly solved to evaluate the impact of the heterogeneous strategies regarding both the power-law exponent and the prefactor in the order-sign autocorrelation function. Our theoretical formulas imply that (i) the power-law exponent formula $\gamma=\alpha-1$ robustly holds even in the presence of the heterogeneous intensity distributions. On the other hand, (ii) the prefactor formula is sensitive to the underlying microscopic assumptions. Indeed, the formula explicitly depends on the intensity distributions among the power-law splitting traders. Furthermore, we find that (iii) the prefactor formula for the homogeneous LMF model systematically underestimates the actual prefactor in the presence of the heterogeneous intensity distributions. We believe that points (i)-(iii) are essential in examining the LMF model for data calibration.

	These days, the availability of high-quality microscopic datasets has been significantly enhanced, and our recent articles~\cite{SatoPRL2023,SatoPRR2023} have verified the LMF prediction quantitatively. Considering such updates from the data-analytic side, we believe that the classical LMF theory should be updated for precise empirical validation. 

	We must admit that our generalisation is just a first step forward for data calibration, and there is plenty of room to improve the trader model for market-order submissions. While only the heterogeneity of the order-splitting strategies is included in our generalised LMF model, other characters, such as the trend-following (herding) behaviour among traders, are not included. Trend-following behaviour is empirically observed at the level of individual traders~\cite{Sueshige2018}, which can be included in the market-order submission models for a more precise market description.

\begin{acknowledgement}
	YS was supported by JST SPRING (Grant Number JPMJSP2110). KK was supported by JST PRESTO (Grant Number JPMJPR20M2), JSPS KAKENHI (Grant Numbers 21H01560, 22H01141, and 23H00467), and JSPS Core-to-Core Program (Grant Number JPJSCCA20200001). 
	
	We thank Hideki Takayasu for his fruitful comment regarding the robustness of our power-law ACF formulas, particularly for the time inhomogeneity of the intensities. We also thank Fabrizio Lillo for his suggestion to study another alternative scenario to explain the long-range correlation, which is based on the superposition of exponential splitters.

	Finally, we declare the author contributions. Both YS and KK contributed to all the analytical calculations. YS initially derived the asymptotic solutions of the heterogeneous LMF model and wrote the programming code. KK followed, confirmed, and fixed all the analytical calculations regarding the mathematical exactness and contributed to the ACF coefficient inequality proofs. Both YS and KK wrote the manuscript and agreed all its contents.

\end{acknowledgement}

\appendix

\section{Master equation for the remaining metaorder length}\label{app:sec:der_ME_remainingMetaorderLength}
	Here we derive the master equation for the remaining metaorder length $R^{(i)}_t$. Let us introduce the arbitrary function $f(R^{i}_t)$ and its dynamics:
	\begin{align}
		f(R^{(i)}_{t+1})-f(R^{(i)}_t) &= 
		\begin{cases}
			0 & \mbox{with prob. $1-\lambda^{(i)}$}\\
			f(R^{(i)}_t-1) - f(R^{(i)}_t) & \mbox{with prob. $\lambda^{(i)}$ when $R^{(i)}_t>1$} \\
			f(L) - f(1) & \mbox{with prob. $\lambda^{(i)}$ when $R^{(i)}_t=1$}
		\end{cases}, \label{app:eq:DynamicsfR}
	\end{align}
	where the renewed metaorder length $L$ obeys $\rho^{(i)}(L)$. By taking the ensemble average of the both-hand sides of \eqref{app:eq:DynamicsfR}, we obtain an identity
	\begin{align}
		&\sum_{R^{(i)}}\left(P_{t+1}(R^{(i)})-P_{t}(R^{(i)})\right)f(R^{(i)}) \\
		=&\sum_{R^{(i)}}P_t(R^{(i)})
		\left[\lambda^{(i)}\left\{ \left(f(R^{(i)}-1) - f(R^{(i)})\right)\Theta(R^{(i)}-1) + \sum_{L=1}^{\infty}\rho^{(i)}(L)\delta_{R^{(i)},1}\left( f(L) - f(1)  \right)
		\right\} 
		\right]\nonumber
	\end{align}
	with the Heaviside function $\Theta(x)$ defined by 
	\begin{equation}
		\Theta(x) = \begin{cases}
			1 & (x > 0) \\
			0 & (x \leq 0)
		\end{cases}.
	\end{equation}
	By setting $f(R^{(i)}_t) = \delta_{R^{(i)}_t,R}$ with a real number $R$, we obtain the master equation for $R>0$ as 
	\begin{align}
		\Delta_{t}P_t(R) = \lambda^{(i)}\left\{ P_t(R+1)-P_t(R) + P_t(1)\rho^{(i)}(R)  \right\},\label{app:eq:FP_RDynamics}
	\end{align}
	where $\Delta_{t}P_t(R) := P_{t+1}(R)-P_t(R)$. This is equivalent to Eq.~\eqref{eq:FP_RDynamics} by replacing the argument dummy variable $R$ with $R^{(i)}$. 

\section{Master equation for the discrete-time Poisson counting process}\label{app:sec:ME_PoissonCountingProcess}
	Let us derive the master equation for the Poisson counting process $N_{t}^{(i)}$. For an arbitrary function $f(N_{t}^{(i)})$, its time-evolution is given by 
	\begin{equation}
		\Delta_t f(N_{t}^{(i)}) = \begin{cases}
			0 & \mbox{with prob. }1-\lambda^{(i)} \\
			f(N_{t}^{(i)}+1) - f(N_{t}^{(i)}) & \mbox{with prob. }\lambda^{(i)}
		\end{cases}
	\end{equation}
	with $\Delta_t f(N_{t}^{(i)}):= f(N_{t+1}^{(i)})-f(N_{t}^{(i)})$. We take the ensemble average of both-hand sides to obtain 
	\begin{align}
		\sum_{N^{(i)}} (P_{t+1}(N^{(i)})-P_{t}(N^{(i)}))f(N^{(i)}) = 
		\sum_{N^{(i)}} \lambda^{(i)} P_{t}(N^{(i)}) \left(f(N^{(i)}+1) - f(N^{(i)})\right).
	\end{align}
	By setting $f(N^{(i)})=\delta_{N^{(i)},N^*}$ with a real number $N^*$ and replacing the dummy variable $N^*$ with $N^{(i)}$, we obtain the master equation
	\begin{equation}
		\Delta_t P_t(N^{(i)}) = \lambda^{(i)}\left(P_t(N^{(i)}-1)-P_t(N^{(i)})\right)
	\end{equation}
	with $\Delta_t P_t(N^{(i)}):=P_{t+1}(N^{(i)})-P_t(N^{(i)})$. Equation~\eqref{eq:P_t(N)_exactsol} satisfies this master equation with the initial condition $P_{t=1}(N^{(i)})=\delta_{N^{(i)},1}$. 
	
\section{Derivation of the exact ACF formula~\eqref{eq:ACF_exact_expon} for the exponential-splitting traders}\label{sec:app:ACF_exact_expon}
	In this Appendix, we exactly derive the ACF formula~\eqref{eq:ACF_exact_expon} under the assumption~\eqref{eq:expon_traders_assump}. Using Eq.~\eqref{eq:exact_sol_ACF}, we obtain 
	\begin{equation}
		\frac{C_\tau^{(i)}}{c_R^{(i)}\left(\lambda^{(i)}\right)^2}= \sum_{R=2}^{\tau}\sum_{N=1}^{R-1} \frac{(\tau-1)!}{(N - 1)!(\tau-N)!}
		\left(\lambda^{(i)}\right)^{N - 1}
		\left(1-\lambda^{(i)}\right)^{\tau-N}
		e^{-(R-1)/L^{*(i)}} + \sum_{R=\tau+1}^\infty \rho_{\geq}^{(i)}(R)
	\end{equation}
	by replacing the dummy variables $R_0^{(i)}$ and $N^{(i)}$ with $R$ and $N$, respectively. By exchanging the sums $\sum_{R=2}^{\tau}$ and $\sum_{N=1}^{R-1}$, the first term of the right-hand side can be evaluated as 
	\begin{align}
		&\sum_{R=2}^{\tau}\sum_{N=1}^{R-1} \frac{(\tau-1)!}{(N - 1)!(\tau-N)!}
		\left(\lambda^{(i)}\right)^{N - 1}
		\left(1-\lambda^{(i)}\right)^{\tau-N}
		e^{-(R-1)/L^{*(i)}} \notag \\
		=& \sum_{N=1}^{\tau-1}\frac{(\tau-1)!}{(N - 1)!(\tau-N)!}
		\left(\lambda^{(i)}\right)^{N - 1}
		\left(1-\lambda^{(i)}\right)^{\tau-N}\sum_{R=N+1}^{\tau} 
		e^{-(R-1)/L^{*(i)}} \notag \\ 
		=& \frac{1}{1-e^{-1/L^{*(i)}}}\sum_{N=1}^{\tau-1}\frac{(\tau-1)!}{(N - 1)!(\tau-N)!}
		\left(\lambda^{(i)}\right)^{N - 1}
		\left(1-\lambda^{(i)}\right)^{\tau-N}\left(e^{-N/L^{*(i)}}-e^{-\tau/L^{*(i)}}\right) \notag \\
		=& \frac{1}{1-e^{-1/L^{*(i)}}}\left[\left(1-\lambda^{(i)}+\lambda^{(i)}e^{-1/L^{*(i)}}\right)^{\tau-1}e^{-1/L^{*(i)}}
		-e^{-\tau/L^{*(i)}}\right].
	\end{align}
	The second term is evaluated as 
	\begin{align}
		\sum_{R=\tau+1}^\infty \rho_{\geq}^{(i)}(R) = \frac{e^{-\tau/L^{*(i)}}}{1-e^{-1/L^{*(i)}}}.
	\end{align}
	We thus obtain Eq.~\eqref{eq:ACF_exact_expon} by using $c_R=1-e^{-1/L^{*(i)}}$.

\section{Derivation of the asymptotic ACF formula~\eqref{eq:LMF_powerlaw_multi_gen} for the power-law-splitting traders}\label{sec:app:ACF_asymp_powerlaw}
	We derive the asymptotic ACF formula~\eqref{eq:LMF_powerlaw_multi_gen} under the assumption~\eqref{eq:power-law_length_assump}. For large $\tau \gg 1$, the binomial distribution~\eqref{eq:P_t(N)_exactsol} can be asymptotically approximated as the normal distribution due to the central limit theorem:
	\begin{equation}
		\lim_{\tau\to \infty} P\left(\frac{N_{\tau}^{(i)}-1-\lambda(\tau-1)}{\sqrt{\lambda^{(i)}(1-\lambda^{(i)})(\tau-1)}}\leq a\right) = \int_{\infty}^a\frac{dx }{\sqrt{2\pi}}e^{-x^2/2} = \frac{1}{2}\left[\erf\left(\frac{a}{\sqrt{2}}\right)+1\right]
		\label{eq:app:CLT}
	\end{equation}
	with the error function defined by 
	\begin{equation}
		\erf(x) := \frac{2}{\sqrt{\pi}}\int_0^x e^{-z^2}dz.
	\end{equation}
	For large $\tau \gg 1/\lambda^{(i)}$, we approximate the sum in the ACF formula~\eqref{eq:exact_sol_ACF} by an integration, such that 
	\begin{align}
		C_\tau^{(i)} &\approx \frac{c_R^{(i)}\left(\lambda^{(i)}\right)^2}{2} \int_2^{\infty} R^{-\alpha^{(i)}}\left[\erf\left(\frac{R-\lambda^{(i)}\tau}{\sqrt{2\tau \sigma^2}}\right)+1\right] dR
	\end{align}
	with $\sigma^2:=\lambda^{(i)}(1-\lambda^{(i)})$ and $c_R^{(i)}=(\alpha^{(i)}-1)/\alpha^{(i)}$, where we use Eq.~\eqref{eq:app:CLT}. By using the partial integration, we obtain 
	\begin{align}
		&\alpha^{(i)}\frac{C_\tau^{(i)}}{\left(\lambda^{(i)}\right)^2} \simeq  \frac{(\alpha^{(i)}-1)}{2}\int_2^{\infty} R^{-\alpha^{(i)}}\left[\erf\left(\frac{R-\lambda^{(i)}\tau}{\sqrt{2\tau \sigma^2}}\right) + 1\right] dR \notag \\ 
		& = -\left[\frac{1}{2}R^{1-\alpha^{(i)}}\left\{\erf\left(\frac{R-\lambda^{(i)}\tau}{\sqrt{2\tau \sigma^2}}\right) + 1\right\}\right]_2^\infty +  \int_2^{\infty} \frac{dR}{\sqrt{2\pi \tau \sigma^2}}R^{1-\alpha^{(i)}}\exp\left(-\frac{(R-\lambda^{(i)}\tau)^2}{2\tau \sigma^2}\right)\notag \\
		& = 2^{-\alpha^{(i)}}\left\{\erf\left(\frac{2-\lambda^{(i)}\tau}{\sqrt{2\tau \sigma^2}}\right)+1\right\} + \int_{2-\lambda^{(i)}\tau}^\infty \frac{dz}{\sqrt{2\pi\tau \sigma^2}}dz(\lambda^{(i)}\tau+z)^{1-\alpha^{(i)}}e^{-z^2/(2\tau \sigma^2)} \notag \\ 
		& \simeq \int_{-\infty}^\infty \frac{dz}{\sqrt{2\pi\tau \sigma^2}}dz(\lambda^{(i)}\tau+z)^{1-\alpha^{(i)}}e^{-z^2/(2\tau \sigma^2)} \simeq \left(\lambda^{(i)}\tau\right)^{1-\alpha^{(i)}},
	\end{align}
	which leads to Eq.~\eqref{eq:LMF_powerlaw_multi_gen}.

\section{Derivation of an inequality for the prefactor upper bound~\eqref{eq:inequality_SKvsLMF}}\label{sec:app:inequality_2}
	Let us prove the inequality 
	\begin{equation}
		\left(\sum_{i=1}^M x_i\right)^a \geq \sum_{i=1}^M x_i^a
		\label{app:ineq2}
	\end{equation}
	for any nonnegative series $x_i \geq 0$, a positive integer $M \geq 1$, and a real number $a>1$. 

	We prove this inequality by mathematical induction. The inequality trivially holds for $M=1$, and let us start the proof from $M=2$ by defining a function $f(x)=(1+x)^a-x^a$. The derivative of $f(x)$ is positive for $x\geq 0$, such that 
	\begin{equation}
		\frac{df(x)}{dx} = a\left\{(1+x)^{a-1}-x^{a-1}\right\} > 0
	\end{equation} 
	since $a-1>0$. We thus find that $f(x)\geq f(0)=1$, or equivalently $(1+x)^a \geq 1 + x^a$. By setting $x=x_2/x_1$ by assuming $x_1 \neq 0$, we obtain $(x_1+x_2)^a \geq x_1^a + x_2^a$. When $x_1=0$, we trivially obtain $(x_1+x_2)^a = x_1^a + x_2^a$. We therefore find 
	\begin{equation}
		(x_1+x_2)^a \geq x_1^a + x_2^a,
		\label{app:ineq3}
	\end{equation}
	which is the special case of the inequality~\eqref{app:ineq2} for $M=2$.

	We next assume that the inequality~\eqref{app:ineq2} holds up to $M=k$ with an integer $k\geq 2$. 
	\begin{equation}
		\left(\sum_{i=1}^{k+1}x_i\right)^a = \left(x_{k+1}+\sum_{i=1}^{k}x_i\right)^a \geq \left(\sum_{i=1}^{k}x_i\right)^a + x_{k+1}^a
	\end{equation}
	by applying the inequality~\eqref{app:ineq3}. Since the inequality~\eqref{app:ineq2} holds for $M=k$, we obtain 
	\begin{equation}
		\left(\sum_{i=1}^{k}x_i\right)^a + x_{k+1}^a \geq \sum_{i=1}^{k} x_i^a + x_{k+1}^a,
	\end{equation}
	which implies that the inequality~\eqref{app:ineq2} holds for $M=k+1$. Thus, the inequality~\eqref{app:ineq2} holds for any integer $M$.

\section{Derivation of the power-law ACF formula~\eqref{eq:superposition-RunDist} as the superposition of exponential splitters with power-law metaorder decay lengths}\label{sec:app:ACF-Ex-Superposition}
	We derive the ACF formula~\eqref{eq:superposition-RunDist} as the superposition of exponential splitting traders. By assuming the approximation~\eqref{eq:approx_noOverwhelmingET}, the formula~\eqref{eq:ACF_exact_expon} is approximately given by 
	\begin{equation}
		C_{\tau}^{(i)} \simeq \left(\lambda^{(i)}\right)^2 e^{- \lambda^{(i)}\tau/L^{*(i)}}
	\end{equation}
	with large $L^{*(i)} \gg 1$ at leading order. Using this approximate formula, we obtain 
	\begin{align}
		C_{\tau} &\simeq \sum_{i\in \Omega_{\rm ET}} \left(\lambda^{(i)}\right)^2 e^{- \lambda^{(i)}\tau/L^{*(i)}} \notag \\
		&= M_{\rm ET}\int_0^1 d\lambda\int_1^\infty dL^*  \left(\frac{1}{M_{\rm ET}}\sum_{i\in \Omega_{\rm ET}} \delta(L^*-L^{*(i)})\delta(\lambda-\lambda^{(i)})\right) \lambda^2 e^{- \lambda\tau/L^{*}} \notag \\
		&= M_{\rm ET} \int_0^1 d\lambda\int_1^\infty dL^*   P_{\rm ET}(L^*,\lambda)\lambda^2 e^{-\lambda \tau/L^{*}},
	\end{align}
	where we used the definition~\eqref{eq:def_continuousP_ET} for $P_{\rm ET}(L^*,\lambda)$.
	By assuming the factorised PDF $P_{\rm ET}(L^*,\lambda)=P_{\rm ET}(L^*)P_{\rm ET}(\lambda)$, we consider the case with a power-law PDF regarding the decay-length $L^{*}$, such that 
	\begin{align}
		P_{\rm ET}(L^*) \simeq (\vartheta-1) \left(L^{*}\right)^{-\vartheta}\>\>\mbox{for $L^{*}\in[1,\infty)$ with $\vartheta \in (1,2)$}.
	\end{align}
	We thus obtain 
	\begin{align}
		C_{\tau} &\simeq (\vartheta-1)M_{\rm ET} \int_0^1 d\lambda P_{\rm ET}(\lambda)\lambda^2 \int_1^\infty dL^* \left(L^*\right)^{-\vartheta} e^{-\lambda \tau/L^{*}} \notag \\
		& = M_{\rm ET}\frac{\vartheta-1}{\tau^{\vartheta-1}}\int_0^1 d\lambda P_{\rm ET}(\lambda)\lambda^{3-\vartheta} \int_{0}^{\lambda \tau} y^{\vartheta-2}e^{-y}dy
	\end{align}
	with the dummy-variable transformation $y:= \lambda \tau / L$. For large $\tau\gg 1$, we asymptotically obtain
	\begin{equation}
		C_{\tau} \simeq \frac{M_{\rm ET}\Gamma(\vartheta)}{\tau^{\vartheta-1}} \int_0^1 d\lambda P_{\rm ET}(\lambda)\lambda^{3-\vartheta}
		= \frac{\Gamma(\vartheta)}{\tau^{\vartheta-1}} \sum_{i\in \Omega_{\rm ET}}\left(\lambda^{(i)}\right)^{3-\vartheta}
	\end{equation}
	with $P_{\rm ET}(\lambda)=(1/M_{\rm ET})\sum_{i\in \Omega_{\rm ET}}\delta(\lambda-\lambda^{(i)})$. This is equivalent to Eq.~\eqref{eq:superposition-RunDist_1}

	We next evaluate the aggregated metaorder-length PDF. Let us use an appoximation $\rho^{(i)}(L)\simeq (1/L^{*(i)})e^{-L/L^{*(i)}}$ to obtain 
	\begin{equation}
		\la L_i\ra \simeq \int_{0}^\infty  \frac{L}{L^{*(i)}}e^{-L/L^{*(i)}}dL = L^{*(i)}. 
	\end{equation}
	Using Eq.~\eqref{eq:MarketAlpha}, we obtain 
	\begin{align}
		\rho_{\rm ST}^{\rm empirical}(L) &\simeq \frac{\la L\ra}{\mu}\sum_{i\in \Omega_{\rm ET}}\frac{\lambda^{(i)}}{\la L_i\ra}\rho^{(i)}(L) \notag \\
		&\propto \sum_{i\in \Omega_{\rm ET}}\lambda^{(i)}\frac{e^{-L/L^{*(i)}}}{\left(L^{*(i)}\right)^2}  \notag \\
		& = \int_0^1 \lambda P(\lambda) d\lambda \int_1^\infty dL^* P(L^*) \frac{e^{-L/L^*}}{L^{*2}} \notag \\
		& \propto L^{-\vartheta-1} \>\>\> \mbox{for large }L \gg 1,
	\end{align}
	implying Eq.~\eqref{eq:superposition-RunDist_2}.

	\section{Numerical LMF simulation method of the heterogeneous exponential splitters}\label{sec:app:SimMethod}
		\begin{figure}
			\centering
			\includegraphics[width=150mm]{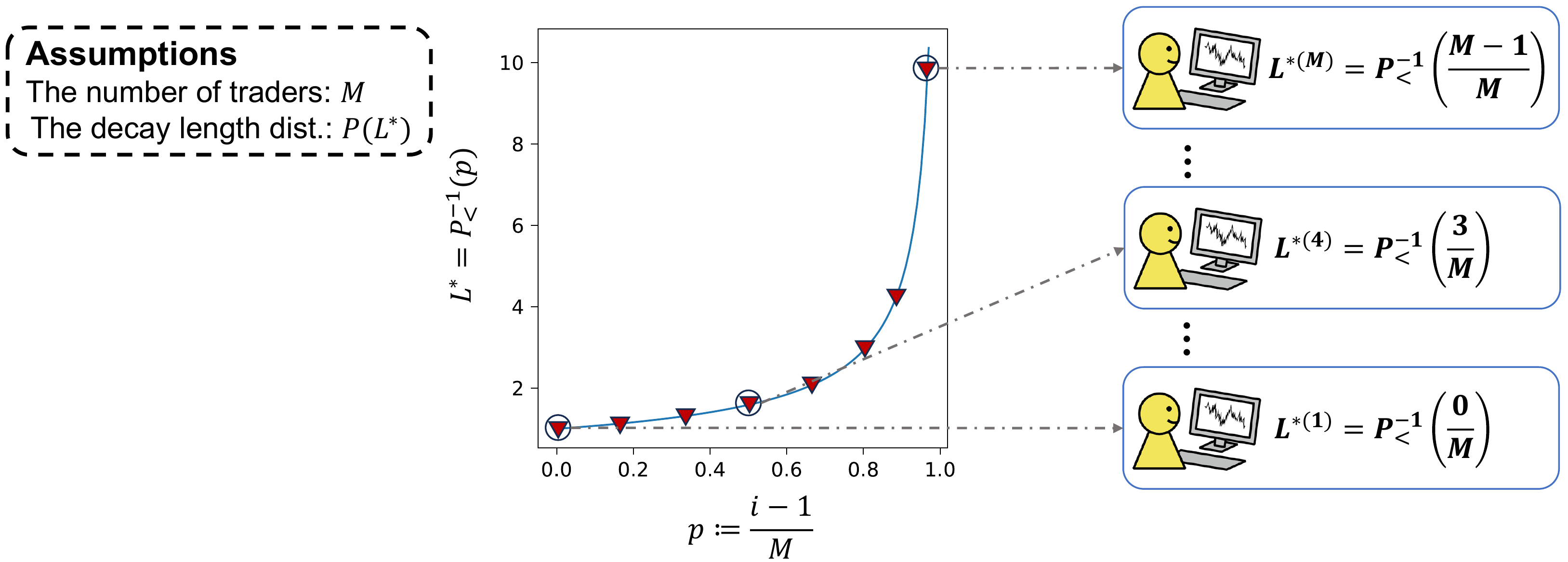}
			\caption{
					A schematic of the systematic allocation of the parameters $\{L^{*(i)}\}_{i \in \Omega_{\rm ET}}$ for $M_{\rm ET}=M$. The characteristic decay lengths $\{L^{*(i)}\}_{i \in \Omega_{\rm ET}}$ are determined by the inverse transform method, such that the empirical PDF $P(L^*):=(1/M_{\rm ET})\sum_{i\in \Omega_{\rm ET}} \delta(L^*-L^{*(i)})$ satisfies the relation~\eqref{eq:app:approx_powerlaw_allocate_L^*}. 
				}\label{fig:schparam}
		\end{figure}
		This Appendix describes the numerical simulation method for the heterogeneous exponential splitting traders, particularly for the deterministic allocation of the parameter $L^*$ based on the {\it inverse transform method} (see Fig.~\ref{fig:schparam} for its schematic). 

		Let us assume that the distribution of $L^*$ is approximately given by 
		\begin{equation}
			P(L^*):=\frac{1}{M_{\rm ET}}\sum_{i\in \Omega_{\rm ET}} \delta(L^*-L^{*(i)}) \simeq (\vartheta-1) \left(L^{*(i)}\right)^{-\vartheta}, \>\>\> M_{\rm ET}:=|\Omega_{\rm ET}|
			\label{eq:app:approx_powerlaw_allocate_L^*}
		\end{equation}
		with $\vartheta \in (1,2)$. The aim of this Appendix is to develop a systematic method to deterministically allocate $\{L^{*(i)}\}_{i \in \Omega_{\rm ET}}$ even for finite $M_{\rm ET}$ to approximately satisfy the relation~\eqref{eq:app:approx_powerlaw_allocate_L^*}. Since its CCDF should obey 
		\begin{equation}
			P_{\geq}(L^*) \simeq \left(L^{*(i)}\right)^{-\vartheta+1}, 
		\end{equation}
		the value of $L^*$ corresponding to the lower-$p \times 100$ percentile is given by $1-p=P_{\geq}(L^*)\Longleftrightarrow L^*=P^{-1}_{\geq}\left(1-p\right)$, where the inverse function is given by 
		\begin{align}
			P^{-1}_{\geq}\left(1-p\right):= \left(\frac{1}{1-p}\right)^{\frac{1}{\vartheta-1}}.
		\end{align}
		On the basis of this relationship, we allocate the parameter $L^{*(\ID)}$ for the trader $\ID$ by 
		\begin{align}
			L^{*(\ID)}= 
				\left\{
					\frac{1}{1-\frac{(\ID-1)}{M_{\rm ET}}}
				\right\}^{\frac{1}{\vartheta-1}}.
		\end{align}

\section{The power-law ACF formula as the superposition of exponential splitters with power-law intensity distribution}\label{sec:app:ACF-Ex-Superposition-Intensity}
	In this Appendix, we study an alternative theoretical scenario for the origin of the LRC as the superposition of exponential splitters with various trading speeds. 

\begin{figure}
		\centering
		\includegraphics[width=150mm]{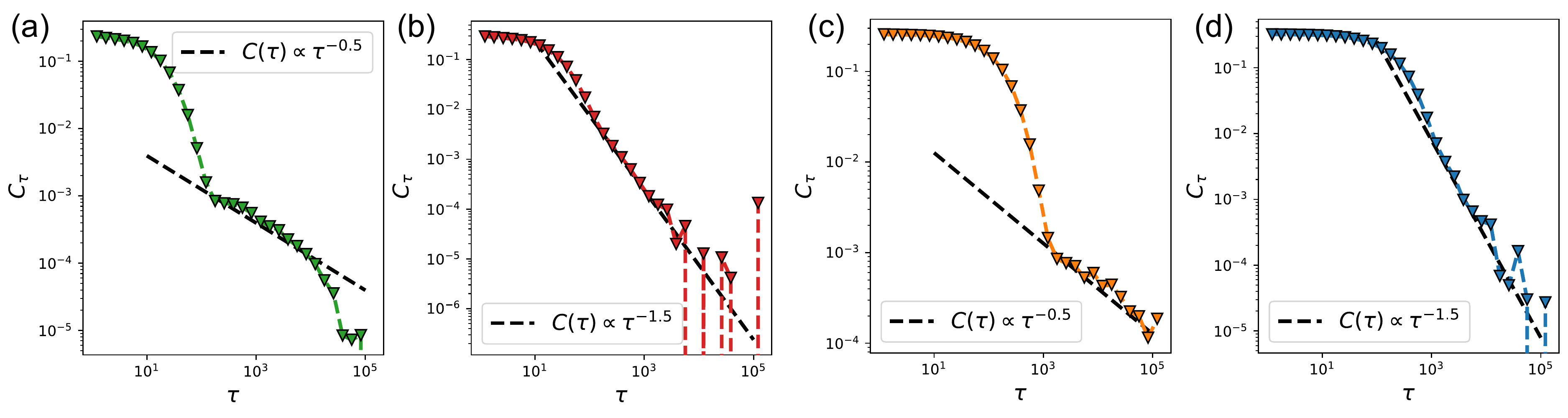}
		\caption{
				Comparisons between the numerical results and the asymptotic theoretical prediction in the system constituted by exponential splitting traders with various intensities. We fixed the number of traders $|\Omega_{\rm EX}|$ with $10^3$ in this simulation.
				(a)~The autocorrelation function of the order flow under $P(\lambda^{(i)})=(\lambda^{(i)})^{-1.5}$, and $L^{*(i)}=10.0$.
				(b)~The autocorrelation function of the order flow under $P(\lambda^{(i)})=(\lambda^{(i)})^{-0.5}$, and $L^{*(i)}=10.0$.
				(c)~The autocorrelation function of the order flow under $P(\lambda^{(i)})=(\lambda^{(i)})^{-1.5}$, and $L^{*(i)}=100.0$.
				(d)~The autocorrelation function of the order flow under $P(\lambda^{(i)})=(\lambda^{(i)})^{-0.5}$, and $L^{*(i)}=100.0$.}
		\label{fig:LMFSuperPosition-Intensity}
	\end{figure}

	Let us assume that there is no dominantly frequent order-splitting trader such that $\lambda_i \ll 1$ for all $i\in \Omega_{\rm ET}$ and that the empirical PDF for the characteristic constants $P_{\rm ET}(L^*,\lambda)$ is factorised such that $P_{\rm ET}(L^*,\lambda)=P_{\rm ET}(L^*)P_{\rm ET}(\lambda)$. Particularly, we focus on the case with a truncated power-law intensity PDF, such that 
	\begin{align}
		P_{\rm ET}(\lambda) \simeq \frac{\beta}{\lambda_{\cut}^{-\beta}-1} \lambda^{-\beta-1} \>\>\>\mbox{for }\lambda\in[\lambda_{\cut},1],
	\end{align}
	where $\lambda_{\cut}$ is a nonzero small parameter representing the lower cutoff of the intensities. We thus obtain 
	\begin{align}
		C_{\tau} &\simeq \frac{M_{\rm ET}\beta}{\lambda_{\cut}^{-\beta}-1}  \int_1^\infty dL^* P(L^{*}) \int_{\lambda_{\cut}}^1 d\lambda \lambda^{-\beta-1}\lambda^2  e^{-\lambda \tau/L^{*}} \notag \\
		&= \frac{M_{\rm ET}\beta}{\lambda_{\cut}^{-\beta}-1} \tau^{\beta-2} \int_1^\infty \left(L^*\right)^{2-\beta}P(L^{*})dL^*  \int_{\lambda_{\cut}\tau / L^*}^{\tau/L^*}  x^{1-\beta}  e^{-x}dx \notag \\
		&= \frac{\beta}{\lambda_{\cut}^{-\beta}-1} M_{\rm ET} \tau^{-(2-\beta)} \int_1^\infty \left(L^{*}\right)^{2-\beta} P(L^{*}) dL^* 
		\left[\Gamma\left(2-\beta,\frac{\lambda_{\cut}\tau}{L^*}\right)-\Gamma\left(2-\beta,\frac{\tau}{L^*}\right) \right],
	\end{align}
	where we apply the variable transformation $x:= \lambda \tau / L^*$ on the second line. Here we focus on the intermediate asymptotic regime $1\ll \tau \ll \lambda^{-1}_{\cut}$. Since $L^*$ is not smaller than one, we obtain asymptotic relations for the incomplete Gamma function  
	\begin{equation}
		\Gamma\left(2-\beta,\frac{\lambda_{\cut}\tau}{L^*}\right) \simeq \Gamma(2-\beta), \>\>\>  
		\Gamma\left(2-\beta,\frac{\tau}{L^*}\right) \simeq \left(\frac{\tau}{L^*}\right)^{1-\beta}e^{-\tau/L^*}
	\end{equation}
	for $L^*\ll \tau \ll \lambda^{-1}_{\cut}$. 
	We thus obtain the power-law ACF decay independent of the metaorder-length distribution until the cutoff time $\lambda^{-1}_{\cut}$ (see Fig.~\ref{fig:LMFSuperPosition-Intensity} for numerical comparison),
	\begin{screen}
		\begin{align}
			C_\tau &\propto \tau^{-(2-\beta)} \>\>\> \mbox{for }1\ll \tau \ll \lambda^{-1}_{\cut}. 
		\end{align}
	\end{screen}
	Beyond the cutoff time $\tau \gg \lambda^{-1}_{\cut}$, the ACF decay should depend on the details of the metaorder-length distribution. 

\end{document}